\documentclass{jfm}
\usepackage[squaren,Gray,cdot]{SIunits}
\usepackage{amssymb}
\usepackage{amsmath}
\usepackage{graphicx}
\usepackage{cite}
\usepackage{color}
\usepackage{subcaption}
\usepackage{float}
\usepackage{bm}
\usepackage{tabularx}
\usepackage[dvipsnames]{xcolor}

\newcommand{\ve}[1]{\mathbf{#1}}
\newcommand{\bu}{\ve{u}}

\newcommand{\bI}{\ve{I}}
\newcommand{\bWe}{{\rm{We}}\to \infty}
\newcommand{\bO}{\bm{\Omega}}

\begin{document}

\shorttitle{Ligament formation in aerobreakup} 
\shortauthor{B. Dorschner et al. } 

\title{On the formation and recurrent shedding of ligaments in droplet aerobreakup}

\author
 {
  Benedikt Dorschner\aff{1}
  \corresp{\email{bdorschn@caltech.edu}},
  Luc Biasiori-Poulanges\aff{2},
  Kevin Schmidmayer\aff{1},
  Hazem El-Rabii\aff{2}
  \corresp{\email{hazem.elrabii@cnrs.pprime.fr}},
  \and
  Tim Colonius\aff{1}
  }

\affiliation
{
\aff{1}
Division of Engineering and Applied Science, California Institute of Technology,\\ 
1200 E.\ California Blvd., Pasadena, CA 91125, USA
\aff{2}
Institut Pprime, CNRS UPR 3346 -- Universit\'e de Poitiers -- ISAE-ENSMA,\\ 1 avenue Cl\'ement Ader, 86961 Futuroscope, France
}

\maketitle
\vspace{-0.1cm}
\begin{abstract}
The breakup of water droplets when exposed to high-speed gas flows is investigated using both high-magnification shadowgraphy 
experiments as well as fully three-dimensional numerical simulations, which account for viscous as well as capillary effects.
After thorough validation of the simulations with respect to the experiments, we elucidate the ligament formation process and the effect of surface tension.
By Fourier decomposition of the flow field, we observe the development of specific azimuthal modes, which destabilize the liquid sheet surrounding the droplet. Eventually, the liquid sheet is ruptured, which leads to the formation of ligaments. 
We further observe the ligament formation and shedding to be a recurrent process.
While the first ligament shedding weakly depends on the Weber number, subsequent shedding processes seem to be driven primarily by inertia and the vortex shedding in the wake of the deformed droplet.
\end{abstract}
\vspace{-0.7cm}
\section{Introduction}
\label{introduction}
The interaction of a droplet with a gas stream involves a complex synergy of
aerodynamic forces and hydrodynamic instabilities that results in  
deformation and fragmentation. This phenomenon is referred as droplet
aerobreakup, which occurs naturally during the fall of rain drops, for instance, and
in variety of technical applications including fuel injection
\citep{allison2016quantitative}, pharmaceutical sprays
\citep{bolleddula2010impact}, and explosion hazards
\citep{eckhoff2016explosion}. Over the last decades, the
aerobreakup phenomenology has been addressed extensively using experimental and
numerical diagnostics, providing mostly two-dimensional (2D) data. As a result,
a comprehensive understanding of the three-dimensional (3D) 
droplet fragmentation mechanisms remains elusive
\citep{chen2008flow,meng2018numerical}. In particular, the ligament formation
process and their subsequent breakup is still poorly understood
\citep{jalaal2014transient}.
\\
Early studies of droplet aerobreakup have identified various 
droplet morphologies by varying the flow conditions and droplet fluid
properties and the underlying deformation mechanisms have been classified. For
high density ratios and Reynolds numbers
(${\rm{Re}}=\rho_{g}D_{0}u_{\infty}/\nu_{g}$), \cite{hinze1955fundamentals} first
defined breakup modes and their transition based on the Weber number
(${\rm{We}}=\rho_{g}D_{0}u_{\infty}^{2}/\gamma$) and the Ohnesorge number
(${\rm{Oh}}=\nu_{l}/\sqrt{\rho_{l}D_{0}\gamma}$). Subsequently,
\cite{krzeczkowski1980measurement} proposed to map transitions of the various
breakup regimes on a ${\rm{We}}$-${\rm{Oh}}$ diagram, and by now a large number of
studies have contributed to this map \citep[see
reviews][]{pilchuse,hinze1955fundamentals,faeth1995structure,guildenbecher2009secondary,guildenbecher2011droplet,lefebvre2017atomization}.
Although there is a good agreement on the description of the various
morphologies of the deformed droplet, regime transitions (in terms of
${\rm{We}}$ and ${\rm{Oh}}$) and the mechanisms involved in the breakup process 
have been subject of debate.
\\
Before the past two decades, the prevailing view was that the mode of droplet
breakup can be classified in five regimes for low Ohnesorge numbers (${\rm{Oh}} <0.1$), namely vibrational, bag, multimode, stripping and catastrophic
breakup. The vibrational regime occurs for ${\rm{We}}<11$ due to the unstable
development of oscillations at the natural frequency
\citep{pilchuse,shraiber1996deformation,wierzba1990deformation} of the droplet
causing its breakup into large fragments. Increasing the Weber number up to 80,
the aerobreakup is driven by the Rayleigh-Taylor instability (RTI) and breakup
modes are distinguished by their wavenumber. The one-wave configuration
corresponds to the bag breakup regime
\citep{lane1951shatter,magarvey1956free,fishburn1974boundary,gel1974various,jalaal2012fragmentation,kulkarni2014bag,wang2014modeling}
where the droplet is first deformed into a disc shape and then a thin hollow bag attached to a toroidal rim, which is blown downstream and finally bursts.
Later, the toroidal rim breaks up due to Rayleigh-Plateau instability
\citep{jain2015secondary}. When the wavenumber increases, more complex bag
structures (including stamen
\citep{hanson1963shock,hirahara1992experimental,gelfand1996droplet,zhao2010morphological,zhao2013temporal} and
multiple bags
\citep{krzeczkowski1980measurement,hsiang1992,hsiang1995drop,hsiang1993drop,cao2007new})
are formed and fragmented, following a similar process. These structures are
referred to as the multimode breakup regime. For ${\rm{We}}<350$, capillary forces are
overcome by shear effects and thus the breakup occurs due to the stretching
of ligaments at the droplet periphery. Literature relates two competing modes
for this Weber range known as shear-stripping regime
\citep{simpkins1972water,hsiang1992,ranger1969aerodynamic,
chou1997temporal} and shear-thinning regime
\citep{liu1997analysis,han1999secondary,lee1999modeling,lee2000experimental,lee2001effect,han2001secondary}.
Finally, for ${\rm{We}}>350$, literature reports a highly contested regime called
catastrophic breakup
\citep{harper1972breakup,reinecke1975shock,hwang1996breakup,joseph1999breakup,theofanous2008},
related to the unstable growth of waves on the droplet upstream side (owing
to RTI). It is suggested that the droplet breaks up when the amplitude of
the waves reach the size of the drop.
\\
Recently, significant progress has been made with the experimental work of
\cite{theofanous2004aerobreakup} and \cite{theofanous2008} on the aerobreakup
in rarefied supersonic flows, which was addressed by means of shadowgraphs and
laser-induced fluorescence. In particular, two major advances have been made.
First, while the literature already relates the occurrence of the RTI at
moderate Weber numbers, these authors observed corrugations due to the
Kelvin-Helmholtz instability (KHI) for higher Weber numbers.  Secondly, they
showed that the catastrophic breakup regime is an
artifact due to the line-integrated nature of shadowgraph visualizations of the
3D complex flow field at the upstream area of the droplet. As a result, they
suggested a reclassification of breakup modes based on the hydrodynamic
instabilities driving the aerobreakup. 
Two regimes are then proposed: the Rayleigh-Taylor Piercing (RTP), driven by
RTI, combined with aerodynamic drag forces and the Shear-Induced
Entrainment (SIE), governed by the combined action of the
Kelvin-Helmholtz instabilities, viscous shearing, and local capillary
mechanisms \citep{theofanous2011aerobreakup}. Compared to the previous
classification, the RTP includes bag and multi-modes regimes while the SIE
refers to the sheet-stripping (or sheet-thinning) mode
\citep{meng2016numerical}. The SIE is proposed as the terminal regime for
${\rm{We}}>10^{3}$.
\\
The ligament formation process in the vicinity of the RTP-SIE
transition and beyond (i.e., ${\rm{We}}>10^2$) is, in particular, a subject of current investigation
\citep{jalaal2014transient,jain2015secondary,meng2018numerical,dorschner2019shock}.
This is mostly due to the
large range of spatial and temporal scales combined to the 3D nature of the breakup,
which surpass traditional 2D experimental and numerical diagnostics used
and thus require sophisticated techniques to elucidate the intricate
breakup mechanisms (i.e., 3D simulations or high-magnification and frequency
optical diagnostics). For ${\rm{We}}>10^2$, a liquid sheet is stretched from the
droplet periphery forming a cylindrical liquid curtain around the droplet
body. The axial symmetry of the liquid sheet is perturbed by the
development of instabilities arising at the liquid sheet surface. Due to
these growing instabilities, the liquid sheet is then disintegrated into
ligaments, which are stretched and broken up into smaller droplets. In an
attempt to describe the instabilities arising on the liquid sheet,
\cite{liu1997analysis} proposed the sheet-thinning mechanism, initially
proposed by \cite{samuelsen1990experimental}.
\cite{samuelsen1990experimental} showed that a liquid sheet subject to
coflowing gases results in `cellular breakup patterns'
\citep{samuelsen1990experimental} and subsequently in the formation of
ligaments due to growing streamwise and spanwise vortical waves on the
liquid sheet surface. Considering high-speed gas flows, the streamwise
waves dominate and thus streamwise ligaments are formed. Ultimately,
ligaments break up into droplets. This mechanism is qualitatively supported
by experimental observations
\citep{lee1999modeling,lee2000experimental,lee2001effect}, 2D numerical
simulations \citep{han1999secondary,han2001secondary,wadhwa2007transient}
and 3D volume-of-fluid simulations \citep{khosla2006detailed,
jain2015secondary}. 
Recently, \cite{jalaal2014transient} proposed the transverse azimuthal
modulation concept \citep{marmottant2004spray,kim2006toward} as an alternative
mechanism to describe the instabilities growing on the liquid sheet. The authors
argued that primary Kelvin-Helmholtz (KH) waves may be subjected to a
transverse destabilization owing to RTI. Growing transverse crests on the
Kelvin-Helmholtz waves are dragged with the flow to form ligaments in the
streamwise direction. The authors provided good qualitative agreement supporting
the transverse azimuthal modulation concept by running 3D numerical simulations
of droplet aerobreakup for Weber numbers up to 200. Due to the lack of experimental
observations of such a destabilization, they attempt to compare their numerical
simulations with theoretical predictions but failed to find conclusive
quantitative evidence. The authors suspect the `simplifications in the
current theories' to be responsible for their mismatch.
\\
Most recent studies on the ligament formation are reported by
\cite{meng2018numerical} through 3D numerical simulations. Comparing the
magnitude of the streamwise and spanwise vorticity captured, the authors found poor
agreement with the sheet-thinning mechanism proposed by \cite{liu1997analysis}.
They pointed out a loss of symmetry of the liquid sheet drawn from the
periphery, which could support the azimuthal modulation mechanism proposed by
\cite{jalaal2014transient}. In an attempt to provide quantitative evidence,
they performed an azimuthal Fourier decomposition of the velocity flow field,
which showed only broadband instability growth for all modes and hence did not provide further evidence of transverse RTI.
\\
In this paper, we perform fully three-dimensional numerical simulations of
aerobreakup events for a moderate Weber number and one without accounting for surface tension to elucidate 
the mechanisms responsible for the ligament formation and the role of surface tension.
We also perform matched aerobreakup experiments using a shock-tube
facility. The breakup events are recorded by means of high-magnification,
high-speed shadowgraphy. Quantitative evidence of the azimuthal modulation are found and discussed. Secondly, we report what we believe to be the
first observation of {\it recurrent} shedding of ligaments. The paper is
structured as follows. The numerical model and the experimental set-up are
described in \textsection 2 and \textsection 3, respectively. The validation of the
numerical simulation with respect to the experiments is presented in
\textsection 4. The mechanisms responsible for the formation of ligaments and 
their recurrent shedding behavior are discussed in \textsection 5. 
Finally, concluding remarks are made in \textsection 6.
%
\section{Numerical modeling} \label{sec:numerics}
The numerical simulation of aerobreakup is a computationally demanding task due to 
the broad physics occurring at a large range of spatio-temporal scales. 
In general, aerobreakup is governed by the compressible Navier-Stokes for 
the liquid and surround gas flow, and coupled by continuity and an equality of stresses at the deforming surface.  This can be modeled by coupling two solvers, or, more commonly by adopting a volume-of-fluid approach and either explicitly tracking the interface or by capturing a slightly diffused interface on the grid \citep{fuster2018review}.
Examples of interface-tracking approaches include free-Lagrange methods
\citep{ball2000shock}, level-set/ghost-fluid 
approaches \citep{abgrall2001computations,liu2003ghost,liu2011conservative,pan2018conservative} or
front-tracking schemes \citep{cocchi1997riemann}.
While such interface-tracking approaches have the advantage of a well-defined, 
sharp interface between components and thus (potentially) accurate interface dynamics, 
various issues ranging from spurious pressure oscillations 
near the interface to lack of conversation make these schemes less suitable 
for shock-dominated flows or aerobreakup (see,
e.g., \cite{fuster2018review} for a recent review on the topic).

Hence, to accurately simulate aerobreakup of a water droplet, we resort to  an
interface-capturing scheme, combining a multicomponent flow model with a
shock-capturing finite-volume method.  These schemes are also known as diffuse
interface methods as the interface is not sharp and tracked explicitly but the
scheme permits some numerical diffusion of the interface. 
This allows for discrete conservation, consistent thermodynamics in mixture
cells and dynamically appearing or vanishing interfaces.  In addition,
diffuse interface methods are generally more efficient compared to their
interface-tracking counterpart, which is crucial for  
multi-scale problems such as aerobreakup.

While there exists a variety of multicomponent models, we consider 
immiscible fluids in mechanical equilibrium and use the model of \cite{kapila2001}.
However, to ensure robustness and stability of the scheme, a  pressure-relaxation method 
is used to converge from a pressure-disequilibrium formulation to its equilibrium \citep{relaxjcp}.
While some numerical studies of aerobreakup can be found in the literature
(see, e.g., \cite{liu2019simulation,marcotte2019density,meng2018numerical, liu2018numerical,garrick2016phd,jalaal2014transient}), these typically 
impose artificial symmetries, which prohibit the formation of truly three-dimensional instabilities or
do not account for any viscous or capillary effects, which become important at later stages of the breakup. 
In order to capture these effects, we model surface tension as
proposed in \cite{schmidmayer2017capillary}.

The viscous, non-equilibrium-pressure multicomponent model with surface-tension effects for 
two components reads as 
\begin{gather}
    \label{eq_multiphaseModel}
    {\begin{array}{*{20}{l}}
    {\cfrac{{\partial {\alpha _1}}}{{\partial t}} + \bu \cdot \mathbf \nabla {\alpha _1}} &{ = \mu \left( {{p_1} - {p_2}} \right) ,} \\ 
    {\cfrac{{\partial {\alpha _1}{\rho _1}}}{{\partial t}} + \boldsymbol \nabla \cdot \left( {{\alpha _1}{\rho _1}\bu} \right)} &{ = 0 ,} \\ 
    {\cfrac{{\partial {\alpha _2}{\rho _2}}}{{\partial t}} + \boldsymbol \nabla \cdot \left( {{\alpha _2}{\rho _2}\bu} \right)} &{ = 0 ,} \\ 
    {\cfrac{{\partial \rho \bu}}{{\partial t}} + \boldsymbol \nabla \cdot \left( {\rho \bu \otimes \bu + p \bI + \bO -\bm{\tau} } \right)} &{ = \mathbf 0 ,} \\ 
    {\cfrac{{\partial {\alpha _1}{\rho _1}{e_1}}}{{\partial t}} + \boldsymbol \nabla \cdot \left( {{\alpha _1}{\rho _1}{e_1}\bu} \right) + {\alpha _1}{p_1}\boldsymbol \nabla \cdot \bu} &{ =  - \mu {p_I}\left( {{p_1} - {p_2}} \right)+\alpha_1 \bm{\tau}_1: \nabla \bu ,} \\ 
    {\cfrac{{\partial {\alpha _2}{\rho _2}{e_2}}}{{\partial t}} + \boldsymbol \nabla \cdot \left( {{\alpha _2}{\rho _2}{e_2}\bu} \right) + {\alpha _2}{p_2}\boldsymbol \nabla \cdot \bu} &{ = \mu {p_I}\left( {{p_1} - {p_2}} \right) +\alpha_2 \bm{\tau}_2: \nabla \bu ,} \\ 
    {\cfrac{{\partial c}}{{\partial t}} + \bu \cdot \boldsymbol \nabla c} &{ = 0 ,} 
    \end{array}}
\end{gather}
where $\alpha_k$, $\rho_k$, $e_k$ and $p_k$ indicate the volume fraction, the
density, the internal energy and the pressure of component $k$.
The mixture variables for  density, pressure and velocity are denoted by 
$\rho$, $p$ and $\bu$, respectively and are given by
\begin{gather}
    \rho = \sum_{k=1}^2 \alpha _k \rho_k  \quad \text{and} \quad
    p = \sum_{k=1}^2 \alpha _k p_k.
\end{gather}

The capillary tensor reads
\begin{gather}
    \bm{\Omega} = - \sigma \left( {\| {\boldsymbol \nabla c} \| \bI -
    \frac{{\boldsymbol \nabla c \otimes \boldsymbol \nabla c}}{{\| {\boldsymbol
    \nabla c} \|}}} \right),
\end{gather}
where $\sigma$ is the surface-tension
coefficient and $c$ is a color function.  The viscous stress tensor for the mixture
is given by 
\begin{equation}
\bm{\tau}=2\eta\left(  \frac{1}{2}\left( \nabla \bu + (\nabla \bu)^T\right)- \frac{1}{3}(\nabla \cdot \bu) \bI \right), 
\end{equation}
where $\eta$ is the mixture shear viscosity and 
the viscous stress tensor for component $k$ is denoted by $\bm{\tau}_k$.
The pressure-relaxation coefficient is given by $\mu $ and 
the interfacial pressure is
\begin{equation}
{p_I} = \frac{{{z_2}{p_1} + {z_1}{p_2}}}{{{z_1} + {z_2}}}, 
\end{equation}
where ${z_k} = {\rho _k}{a_k}$ is the acoustic impedance of component $k$. 

Due to $p_1 \neq p_2$ in this model, the total-energy equation
of the mixture is replaced by the internal-energy equation for each component.
Nevertheless, the mixture-total-energy equation of the system can be written
in usual form:
\begin{equation}
    \label{eq_total_energy}
    \cfrac{{\partial \rho E + \varepsilon _\sigma}}{{\partial t}} + \boldsymbol \nabla \cdot \left( {\left( {\rho E + \varepsilon _\sigma + p} \right) \bu + \bO \cdot \bu - \bm{\tau} \cdot \bu} \right) = 0 ,
\end{equation}
where the total energy is 
\begin{gather}
    E = e + \frac{1}{2} \| \bu \|^2,
\end{gather}
and the internal energy is given by
\begin{gather}
    \label{eq_internalEnergy}
    e = \sum_{k=1}^2 Y_k e_k \left( \rho_k , p_k \right).
\end{gather}
The capillary energy reads
\begin{gather}
    {\varepsilon _\sigma } = \sigma \| {\boldsymbol \nabla c} \| .
\end{gather}
Note that equation~\eqref{eq_total_energy} is redundant when solving
internal-energy equations for both components . 
However, it is included to ensure total 
energy conservation also numerically (see \citet{relaxjcp} for further details).

In~\eqref{eq_internalEnergy}, $e_k$ is defined
via an equation of state and $Y_k$ are the mass fractions
\begin{gather}
    Y_k = \frac{\alpha_k \rho_k}{\rho} .
\end{gather}
Here, we consider a two-phase mixture of gas ($g$) and liquid ($l$).
The gas is modeled by the ideal-gas equation of state
\begin{gather}
    p_g = ( \gamma_g - 1) \rho_g e_g ,
\end{gather}
with $\gamma_g = 1.4$. The liquid on the other hand is modeled by the
stiffened-gas equation of state
\begin{gather}
    p_l = ( \gamma_l - 1) \rho_l e_l - 
    \gamma_l \pi_\infty,
\end{gather}
where $\gamma_l = 6.12$ and $\pi_\infty = \unit{3.43 \times 10^8}{\pascal}$
(\cite{coralic2014WENO5, meng2014, meng2018numerical}).

Numerically, this model is solved in three steps, which are outlined below.
First, the hyperbolic non-equilibrium-pressure model is solved by neglecting
surface-tension effects and relaxation terms.
Second, the surface-tension model is solved and finally, in the last step, 
the pressure is relaxed until an equilibrium is reached.
In summary the model is solved with the following steps:
\begin{enumerate}
\item Solve \emph{pressure-disequilibrium model} using a Gudonov-type method. 
At the volume-volume interfaces, the associated Riemann problem is computed
using the HLLC approximate solver.
\item  Solve hyperbolic \emph{surface-tension model}, ensuring momentum and energy conservation.
\item  Infinite \emph{pressure relaxation} 
($\mu \to +\infty$), converging to the thermodynamically 
consistent, mechanical-equilibrium model.
\end{enumerate}

A second-order-accurate MUSCL scheme with two-step time integration is used, 
where the first step is a predictor step for the second and the usual 
piece-wise linear MUSCL reconstruction \citep{toro97} is used for the primitive variables.
In addition, the monotonized central (MC, \cite{van1977MC}) slope limiter is combined with
the THINC interface-sharpening technique to minimize interface diffusion \citep{shyue2014thinc}.

In order to resolve the wide range of spatial and temporal scales of 
shock-fronts and interfaces, an adaptive mesh refinement technique is employed \citep{schmidmayer2019AMR}.
The refinement criteria are based on variations of volume fraction, density, 
pressure and velocity.

This methodology has been validated, verified and tested in previous set-ups and is 
implemented in the open-source code ECOGEN \citep{schmidmayer2019ecogen}.

\begin{figure}
  \centerline{\includegraphics[width=0.9\textwidth]{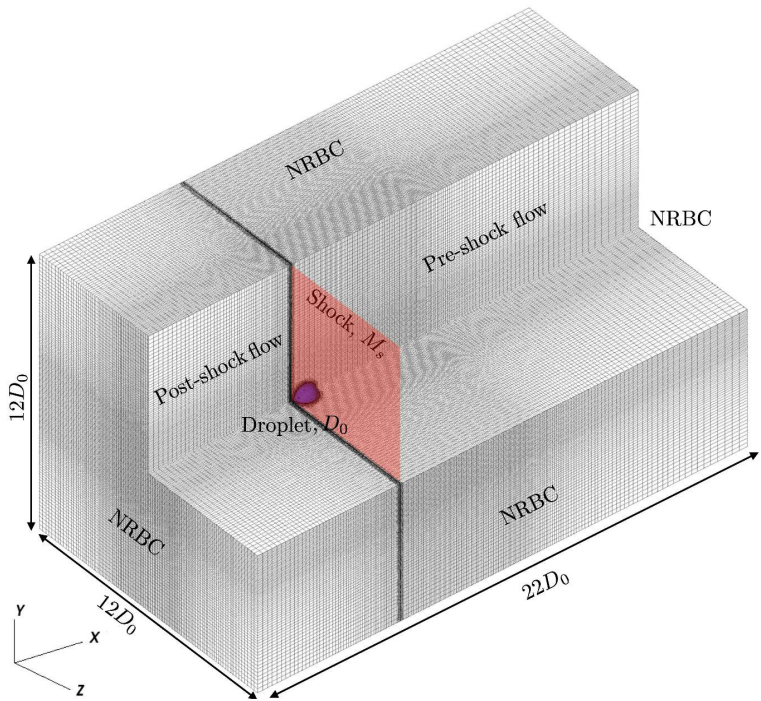}}
  \caption{Initial computational mesh and set-up for the numerical simulation
  of aerobreakup.}
  \label{fig:mesh0}
\end{figure}
\subsection{Problem definition}
Aerobreakup occurs when a liquid drop is suddenly exposed to a high-speed gas
flow. These initial conditions are typically generated using a planar shock
due to its simplicity, robustness and repeatability in both experiments and
simulations without significantly interfering with the droplet or the evolution
of the subsequent stages of the aerobreakup.  In the simulation, we match the
experimentally measured mean shock-strength $M_s \sim1.3$,  the Reynolds number
${\rm{Re}} = \frac{u_s D_0}{\nu}$ and the Weber number ${\rm{We}}=\frac{\rho_g
u_s^2  D_0}{\sigma}$, where $D_0$, $u_s$, $\nu$ and $\sigma$ denote
the initial droplet diameter, the post-shock gas velocity, the kinematic
viscosity of the gas as well as the surface-tension coefficient. In the
experiments we mainly focus on the piercing regime with mean Weber 
numbers in 
the range of ${\rm{We}}=[200, 700]$. Numerically, we conduct two
simulations, where one matches a mean experimental Weber number of ${\rm{We}} \sim 470$ and
one without any surface-tension effects, which is denoted in the following as $\bWe$.
Note that $\bWe$ is purely nomenclature and does not indicate the limiting process to infinite Weber number.
In both cases the Reynolds number
is set to ${\rm{Re}} \sim 7000$. Hence, we numerically probe
and compare both the piercing and the stripping regime.

\begin{figure}
  \centerline{\includegraphics[width=0.6\textwidth]{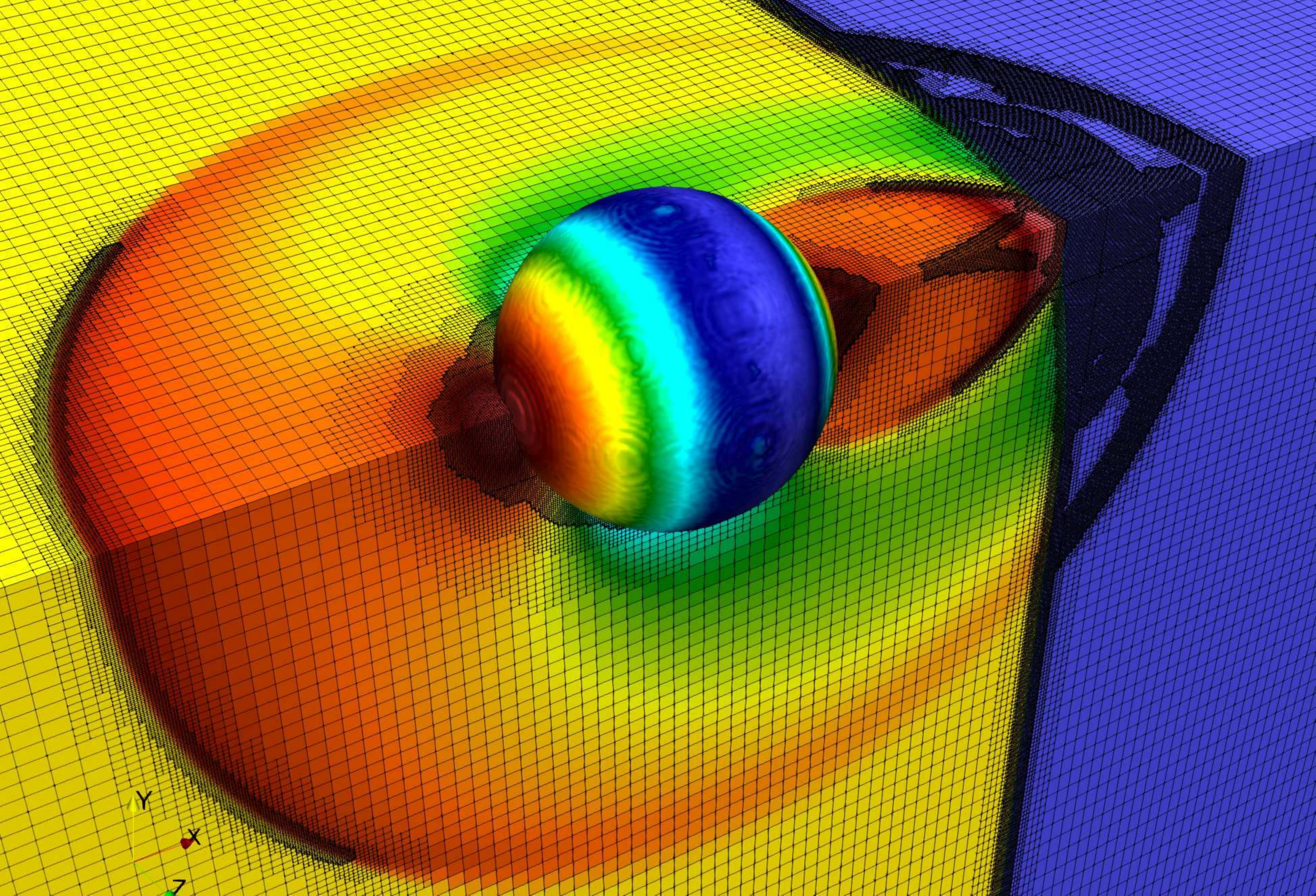}}
  \caption{Snapshot of the adaptively refined mesh, colored by mixture pressure.}
  \label{fig:mesh2f}
\end{figure}

The simulations are carried out in rectangular computational domain, which is
given by $[-7D_0, 15D_0] \times [-6D_0,6D_0]\times [-6D_0,6D_0]$. 
The domain size was chosen based
on sensitivity studies, which aim to both minimize the influence of the
domain boundary conditions as well as the computational effort. Our results are
in line with previous studies \citep{meng2018numerical}.  
To capture the non-axisymmetric, three-dimensional modulation of the droplet and 
its surrounding flow field, we refrain from imposing any symmetries or simplifications and carry 
out full three-dimensional simulations for which the initial computational mesh
and the set-up are shown in figure \ref{fig:mesh0}. The droplet
is initially at rest and placed at the origin. On all domain boundaries,
non-reflective boundary conditions (NRBC) are imposed. 
To ensure high spatial and temporal accuracy at reasonable computational costs,
the adaptive mesh is composed out of four grid levels, which are adapted to
follow the shocks, the interface and the turbulence.
Therefore, as detailed and validated in \cite{schmidmayer2019AMR}, the refinement 
criterion reads 
\begin{equation}
    \frac{|X_{{\rm{Nb}}(i,j) } - X_i|}{\text{min}(X_{{\rm{Nb}}(i,j) } - X_i)} > \varepsilon,
\end{equation}
where  $X$ indicates the flow variable for which we chose
the density, velocity, pressure and volume fraction. The subscript Nb$(i,j)$ denotes the 
neighboring cells of cell $i$.
The threshold is 
conservatively set to $\varepsilon=0.04$.
As a result, the initial droplet is resolved by
$D_p=140$ point per diameter, which was increased compared to
\cite{meng2018numerical} in order to capture capillary effects such as the
formation of ligaments during the course of the breakup. 
Note however, the selected resolution is by no means able to capture all fine-scale 
effects and the resolution required to do so will far exceed 
currently available computational resources, even on the largest of supercomputers (see, e.g., 
\cite{meng2018numerical} for estimates). 
However, as we will demonstrate below in section \textsection \ref{sec:expeirmental_validation}, 
the good agreement with the experiments gives us confidence that most pertinent effects of 
aerobreakup are indeed captured with the selected resolution.
The mesh adaptivity, following both the droplet interface and the shock is exemplified
by a snapshot in figure \ref{fig:mesh2f} during the initial phase of the simulation.
In addition, a
conservative grid stretching towards the domain boundaries as shown in  figure
\ref{fig:mesh0} is used to further aid efficiency of the  computations.
Finally, to avoid spurious symmetries originating from the artificially
symmetric initial conditions, we impose an initial velocity field with
random perturbations of maximum $\mathcal{O}(10^{-4} u_s)$.  For the adaptive
time marching scheme we maintain a maximum Courant-Friedrichs-Lewy (CFL) number
of $0.3$.


\section{Experimental set-up}
\label{sec:experimental_setup}
\begin{figure}
  \centerline{\includegraphics[width=\textwidth]{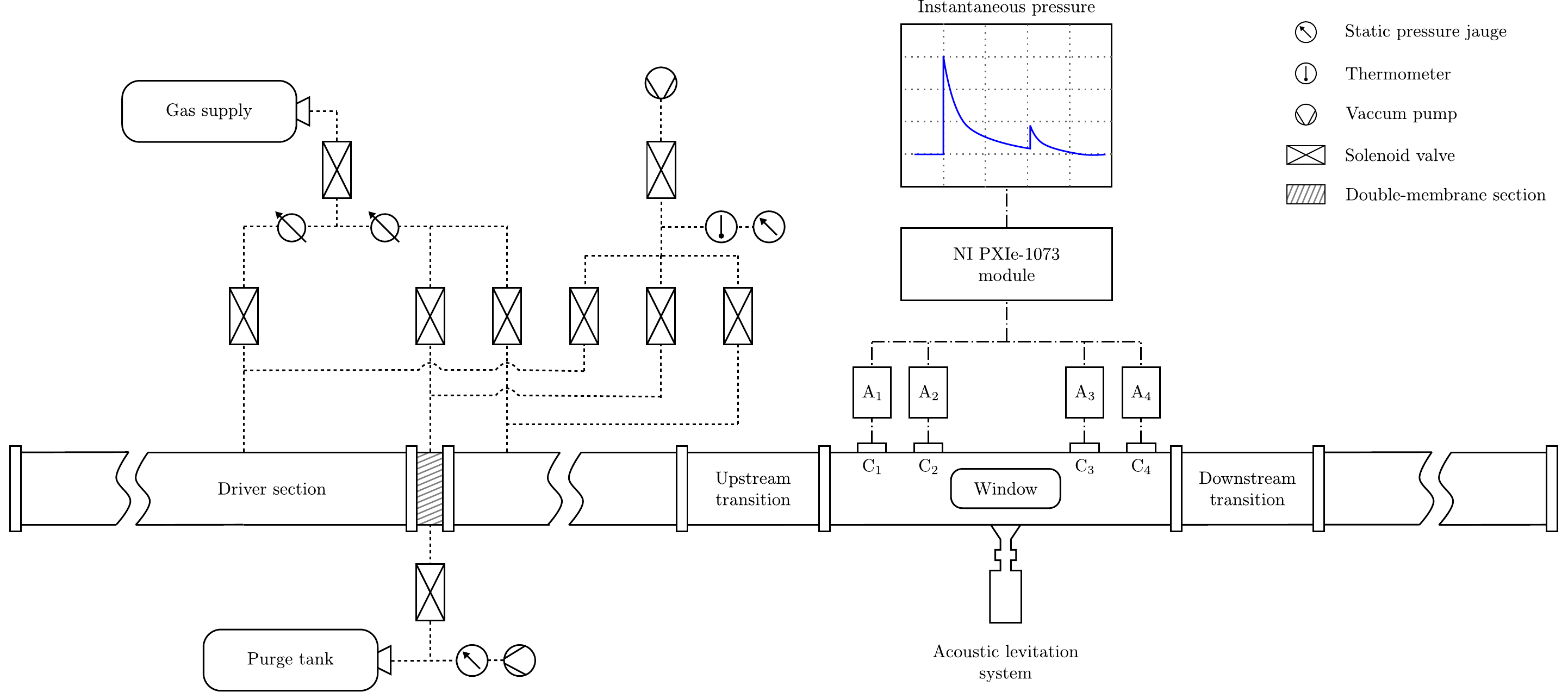}}
  \caption{Shock tube facility (side view).}
  \label{fig:0}
\end{figure}

The shock tube is manufactured from several stainless steel pipes with shell
thickness of $5~\milli\meter$ and circular cross-section of $52~\milli\meter$ in
diameter. The facility consists of four components (Fig.~\ref{fig:0}): a driver
section, a double-membrane section and a driven section which includes a test
section of square cross-section of $46~\milli\meter$ $\times$ $46~\milli\meter$.
Both membranes have the same burst pressure.  The test section is connected on
the driven section by means of circular-to-square transitions. Transitions are
smooth and designed to preserve the area. The upstream transition is placed
$1000~\milli\meter$ ahead from the center of the test section to insure a sharp and planar
front wave profile. The test section is fitted with two oblong BK7 windows
mounted opposite one another on its lateral sides to allow for optical
diagnostics (shadowgraph or laser-sheet visualization).  The double-membrane section and the driver section
are pressurized with air at 75\% and 125\% of the burst pressure of the
membranes (Mylar sheets), respectively.  The shock wave is initiate by abruptly
purging the double-membrane section through an extraction gas port. The driven
section is filled with ambient air at controlled temperature.

To monitor the shock propagation, the instantaneous test section pressure is
measured at four lateral positions by dynamic high-speed pressure transducers (noted C in Fig.~\ref{fig:0})
with acceleration compensation (Kistler 603B). They allow to measure pressure
fluctuations over a range from vaccum to 200 bar~with a rise time of 1~\micro\second. Each sensor is coupled to a Kistler 5018A charge amplifier (noted A in Fig.~\ref{fig:0}) with a
bandwidth of 200~\kilo\hertz~which converts the mechanical stress into an
electrical signal (0-10~\volt). The electrical signal is acquired using a
National Instruments NI PXIe-1073 module with 16 input channels with a
frequency sample of 60~M\hertz. The sensors have an active area of $5~\milli\meter$ diameter and are mounted in Pom-C \textregistered~holders. The
active surface of the sensors is installed flush to the test section wall.
These sensors are used to measure the shock wave velocity and the pressure
jump, and are also exploited for the triggering of the high-speed camera.
Synchronization of the high-speed camera with the light source emission and the
breakup event is performed with a DG535 Digital Delay and Pulse Generator
(Stanford Research system).

During the experiments, the water drop is held in a stable equilibrium at the
center of the test section by the sound radiation pressure of an ultrasonic
standing wave generated by the single-axis acoustic levitator. The levitator consists of a Langevin-type transducer coupled to a
mechanical amplifier with a radiating surface of 35~\milli\meter~in diameter. The
transducer operates at a resonant frequency of 20~\kilo\hertz~and is driven by a 1.5~\kilo\hertz~ultrasonic power supply. The radiating surface is mounted flush with the inside
bottom surface of the chamber. Opposite to it, the upper surface of the chamber
acts as a reflector of the acoustic waves for standing wave generation. To
avoid disturbing the aerobreakup process with the sound radiation pressure, the
levitation system is turned off following a voltage setpoint from a pressure
transducer monitoring the shock propagation.

\begin{figure}
  \centerline{\includegraphics[width=\textwidth]{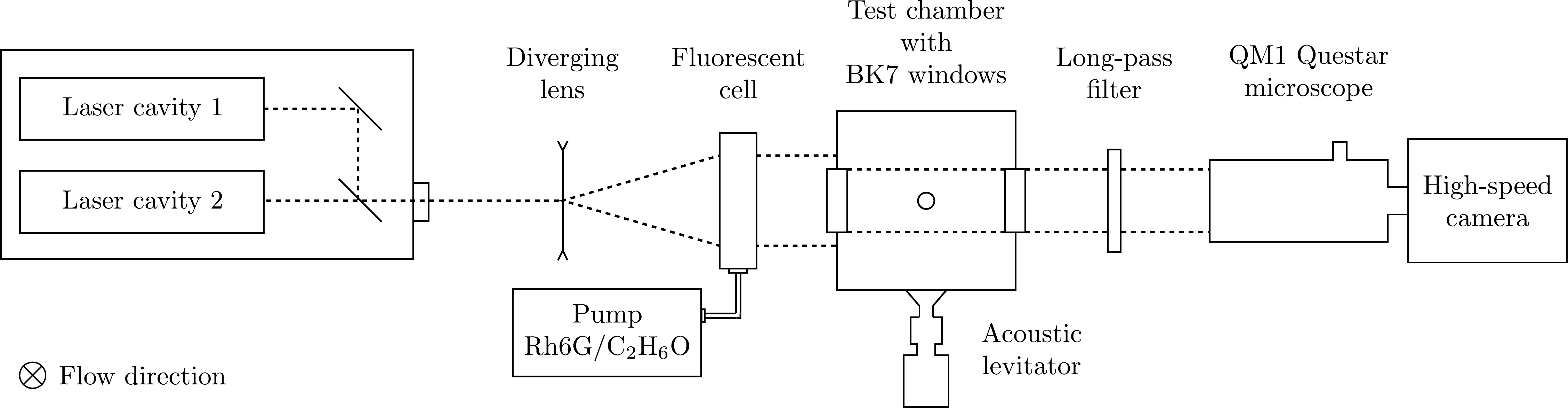}}
  \caption{High-magnification shadowgraphy system (axial view).}
  \label{fig:1}
\end{figure}

Aerobreakup experiments are captured with a high-magnification shadowgraphy
system. The backlight illumination is provided by the laser-induced
fluorescence of a Rhodamine 6G dye solution (Figure \ref{fig:1}). A high-power
dual oscillator/single head diode-pumped Nd:YAG laser (Mesa PIV, Continuum)
delivering short pulses in the $120–180~\nano\second$ range at repetition rate up
to $80~\kilo\hertz$ is used to induce the fluorescence. Visualization is
performed with a high-speed Fastcam Photron SA-Z equipped with a
Maksutov–Cassegrain catadioptric microscope (QM1 Questar). The maximum optical
resolution is 1.6~\micro\meter~with a magnification up to 125:1. The depth of
focus is approximately $0.6~\milli\meter$. The high-speed camera and the
dual-cavity laser are synchronized by a digital delay generator. More details
on the optical diagnostic are available in \cite{biasiori2019high}. Sequence of
images displayed are captured with camera settings adjusted to record frames at
512$\times$904 pixel resolution with a sampling frequency of $40~\kilo\hertz$.
The average laser output power is $30~\watt$ and the pulse width is $174~\nano\second$. The measured spatial resolution of the imaging system was $6.5~\micro\meter$ per pixel. 
The time of the shock-wave interaction with the droplet ($t=0$) is
determined by laser beam deflection with a continuous-wave laser beam
perpendicular to the flow direction and tangent to the droplet front side. The
deflection is recorded with a photodiode with a $0.9~\nano\second$ risetime.

\section{Comparison of simulations and experiments}
\label{sec:expeirmental_validation}


\subsection{Droplet morphology}
In order to compare the numerical simulation with images obtained from shadowgraphy in the following sections,
we extract isosurfaces of
volume fraction $\alpha_l=0.01$ from the simulation, and color them by velocity magnitude (see, also figures \ref{fig:2} and \ref{fig:4}).
Note that in diffuse interface methods, as employed here, the exact location of the interface 
is not well defined and it can only be approximated by isosurfaces of the volume fraction. 
While this leads to an ambiguous representation of the droplet surface, 
a sensitivity study in \cite{meng2018numerical} led to the conclusion that
$\alpha_l=0.01$ was believed to be fair for comparison with experiments, considering the 
obscuring mist of the experiment, generated during the course of the breakup. 
For comparison of the numerical results with the experiment, we use, unless stated otherwise, the following 
non-dimensionalization
\begin{equation}
 \bm{x}^\ast=\frac{\bm{x}}{D_0}, \qquad \tau= t\frac{\bm{u}_s}{D_0} \sqrt{\frac{\rho_s}{\rho}},
\end{equation}
where $\bm{x}$ and $t$ denote the dimensional location and time, respectively.

\subsubsection{Finite Weber number case}
\label{finite_We_case}
\begin{figure}
  \centerline{\includegraphics[width=\textwidth]{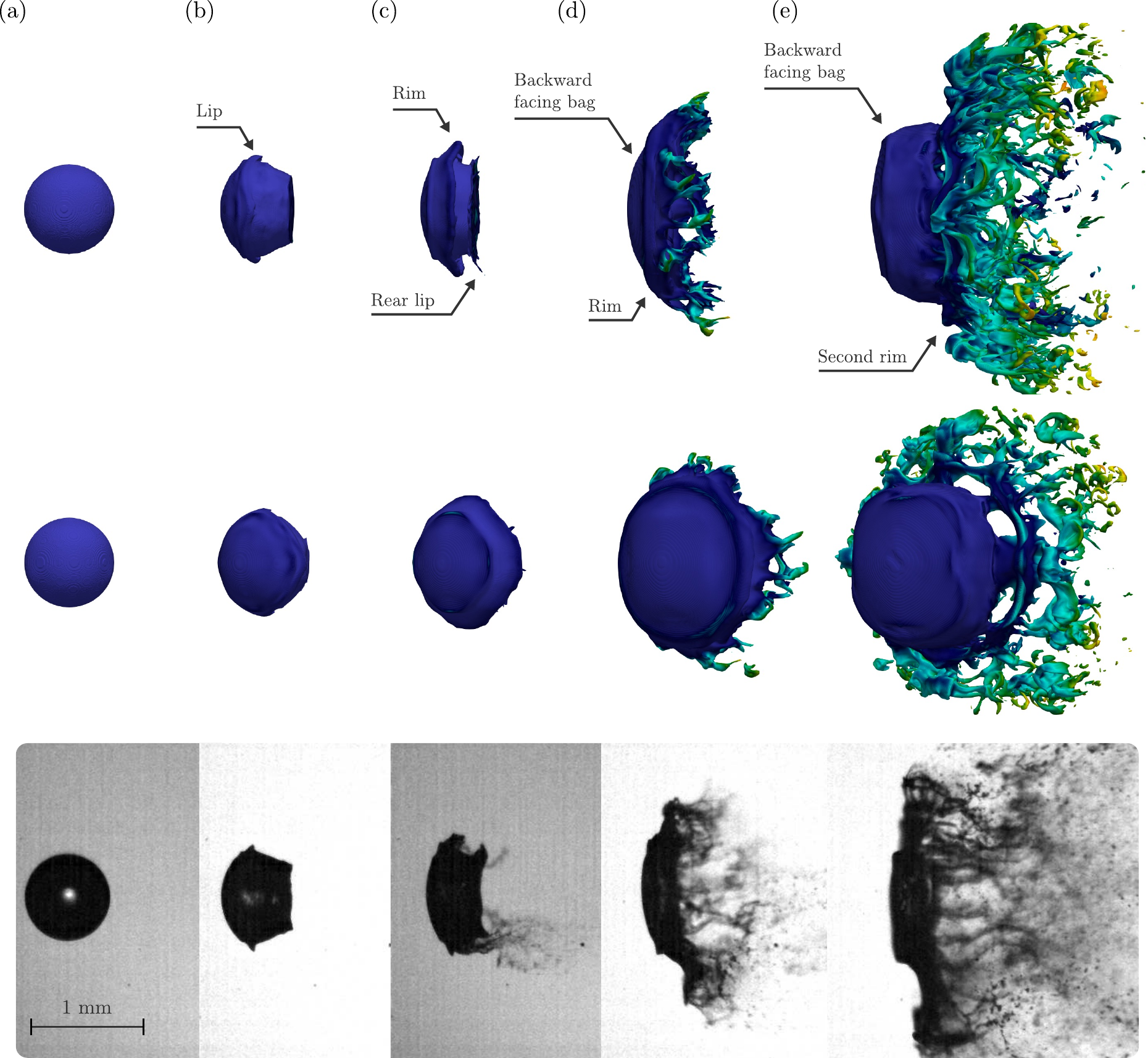}}
  \caption{Numerical simulation of a water droplet aerobreakup (top two rows) and experimental
  visualizations (bottom rows). Characteristic times are (a) 0.00, (b) 0.27, (c) 0.47, (d)
  0.72 and (e) 1.16. The timing information is for both experimental and numerical results expected for the numerical image (e) which corresponds to $\tau$=1.01. For the simulations, an isosurface of volume fraction $\alpha_l=0.01$ colored by velocity magnitude is depicted from two different perspectives.  The scale bar refers to experimental results.  The Weber
  number in the simulation and the experiment are 470 and 492, respectively.}
  \label{fig:2}
\end{figure}

\begin{figure}
  \centerline{\includegraphics[width=\textwidth]{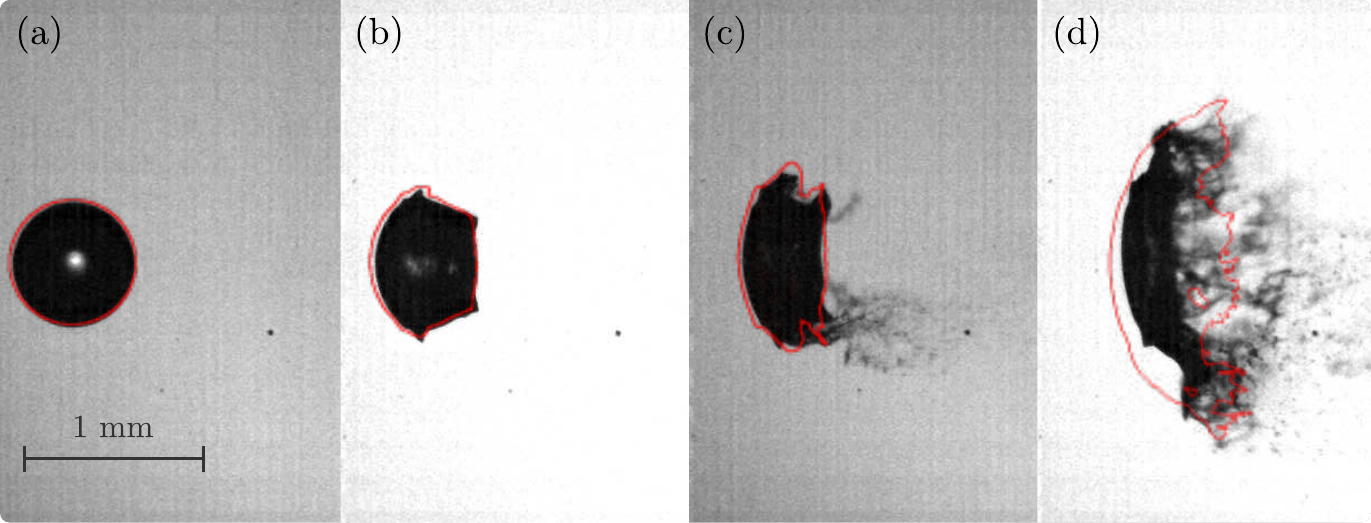}}
  \caption{Overlaying of droplet contours from numerical results (red lines) on
  experimental images at the early stages of the aerobreakup. Characteristic times are (a) 0.00, (b) 0.27, (c) 0.47 and (d) 0.72. The Weber number in the simulation and the experiment are 470 and 492, respectively.}
  \label{fig:3}
\end{figure}

At the first stages of the aerobreakup process ($\tau$ = 0.27 - 0.91),
the numerical results show the typical droplet shapes that experimental
visualizations (Figure \ref{fig:2}) and literature report, with a good
qualitative agreement with our experiment. First, the initial droplet deforms into a muffin-like
shape, described by a spherical upstream side with lips growing in the spanwise
direction. The droplet core takes the shape of a conic cylinder and the
downstream side is flattened into a planar interface ($\tau$ = 0.27). While the
droplet is continuously flattened, the stretching of lips in the streamwise
direction results in a toroidal liquid rim ($\tau$ = 0.47). Simultaneously,
rear lips raise in the downstream side. Owing to inertial forces, the rim at
the droplet periphery is stretched, which deforms the droplet into a
crescent-like shape ($\tau$ = 0.72). The rim begins to disintegrate into ligaments
and subsequently into fragments.
For characteristic times from 0.00 to 0.72,
contours, extracted from the isosurfaces of the volume fraction $\alpha_l=0.01$
of the droplet, are overlaid
on the experimental visualizations in figure \ref{fig:3}. 
The contours
are initially in good agreement with the experimental images. At later stages ($\tau$ = 0.72 - 1.16), the numerical results and the
experimental visualizations both show a periodic distribution of ligaments, though there are discrepancies in the precise shape. 
However, the ligament distribution in the numerical simulation is consistent with the
distribution experimentally observed as detailed below in section \ref{sec:ligaments0}. 
Figure \ref{fig:2} shows that the 
toroidal rim is continuously sheared away with the flow for times $\tau$ = 0.72 - 1.16, 
which leads to a cylindrical curtain surrounding the droplet core and a cavity
behind. This results in a backward facing bag. The downstream end of the bag
bends in the spanwise direction forming a second rim, hindering the flow.
Subsequently, the rim is subject to the development of multiple bags in the
streamwise direction.  Finally, bags are pierced by the flow, similar to bag
and multimode breakup regimes. The ligaments resulting from this piercing process
are tied up to their ends by the annular ring. The periodic nature of the
ligament distribution is further discussed in section \ref{sec:ligaments}.

\subsubsection{Simulation without surface tension}
\label{infinite_We_case}
 
Next, in order to isolate the effect of surface tension during the breakup,
we conduct an additional numerical simulation and set the surface tension to zero with 
$M_{s}$=1.3 and $D_{0}=0.804\milli\meter$ and compare the results with experiments at
a high Weber number of ${\rm{We}}=1100$ ($M_{s}$=1.3, $d_{0}=1.68\milli\meter$). 

\begin{figure}
  \centerline{\includegraphics[width=\textwidth]{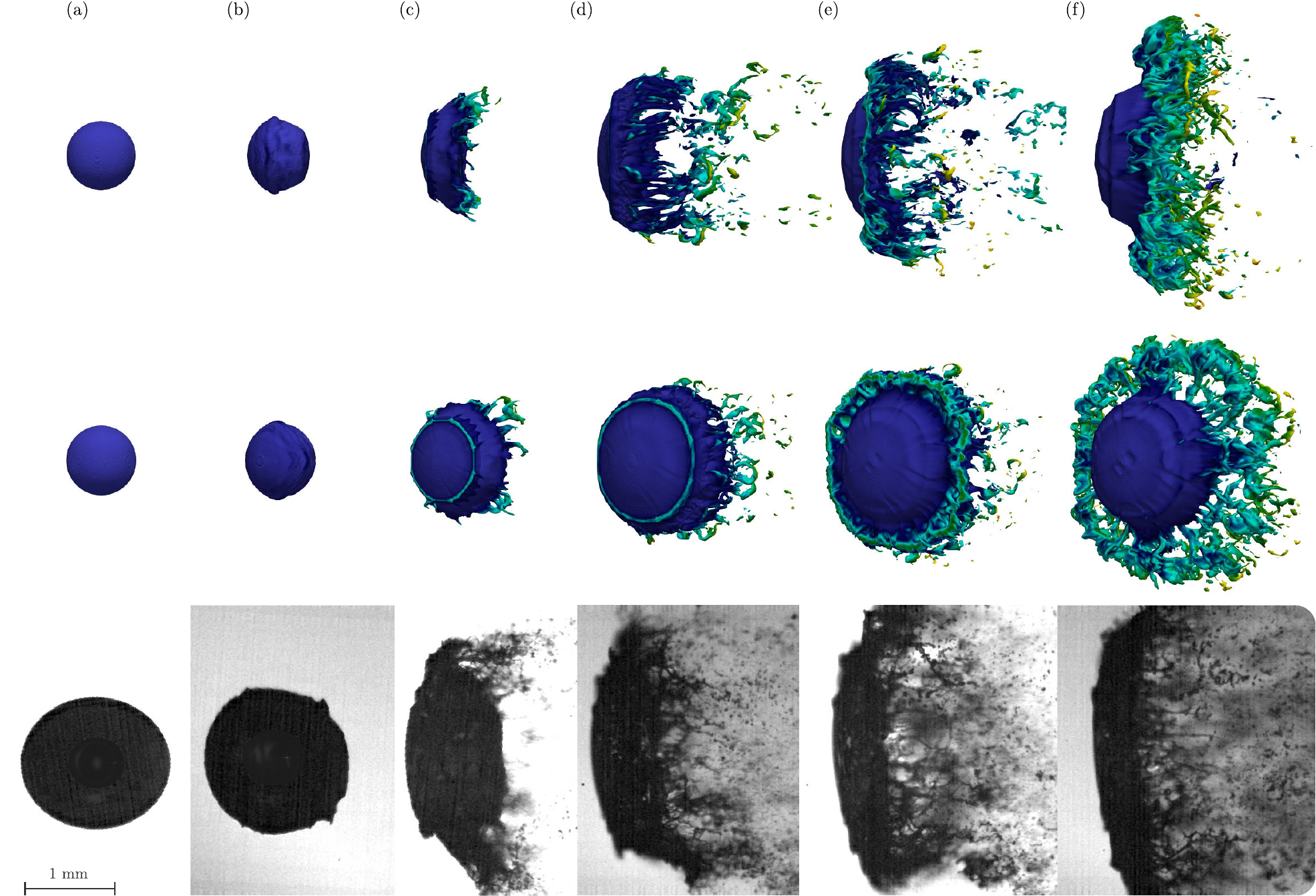}}
  \caption{Numerical simulation of a water droplet aerobreakup and experimental
  visualizations. Characteristic times are (a) 0.00, (b) 0.20, (c) 0.52, (d) 0.73,
  (e) 0.86 and (f) 0.96. The scale bar refers to experimental results. In the experiment, the Weber
  numbers is ${\rm{We}}=1100$, whereas the simulation does not account for surface tension effects, i.e., $\bWe$.}
  \label{fig:4}
\end{figure}

Figure \ref{fig:4} shows that the droplet morphology and the mechanisms
observed bear resemblance to the finite Weber number case, especially 
in the early stages. The droplet is
first deformed into a muffin-like shape with lips growing in the spanwise
direction. Lips are rapidly sheared away with the flow by forming a liquid rim
surrounding the droplet body. The rim stretches in the streamwise direction and
forms a cylindrical curtain, which eventually results in a backward facing bag.
Finally and in contrast to the morphology previously observed in figure
\ref{fig:2}, much less distinct ligaments are formed and the liquid curtain breaks up
directly into fine-scale structures due to a lack of the restoring surface
tension forces.

\subsection{Center-of-mass evolution}
\label{Center-of-mass}
\begin{figure}
  \centerline{\includegraphics[width=0.6\textwidth]{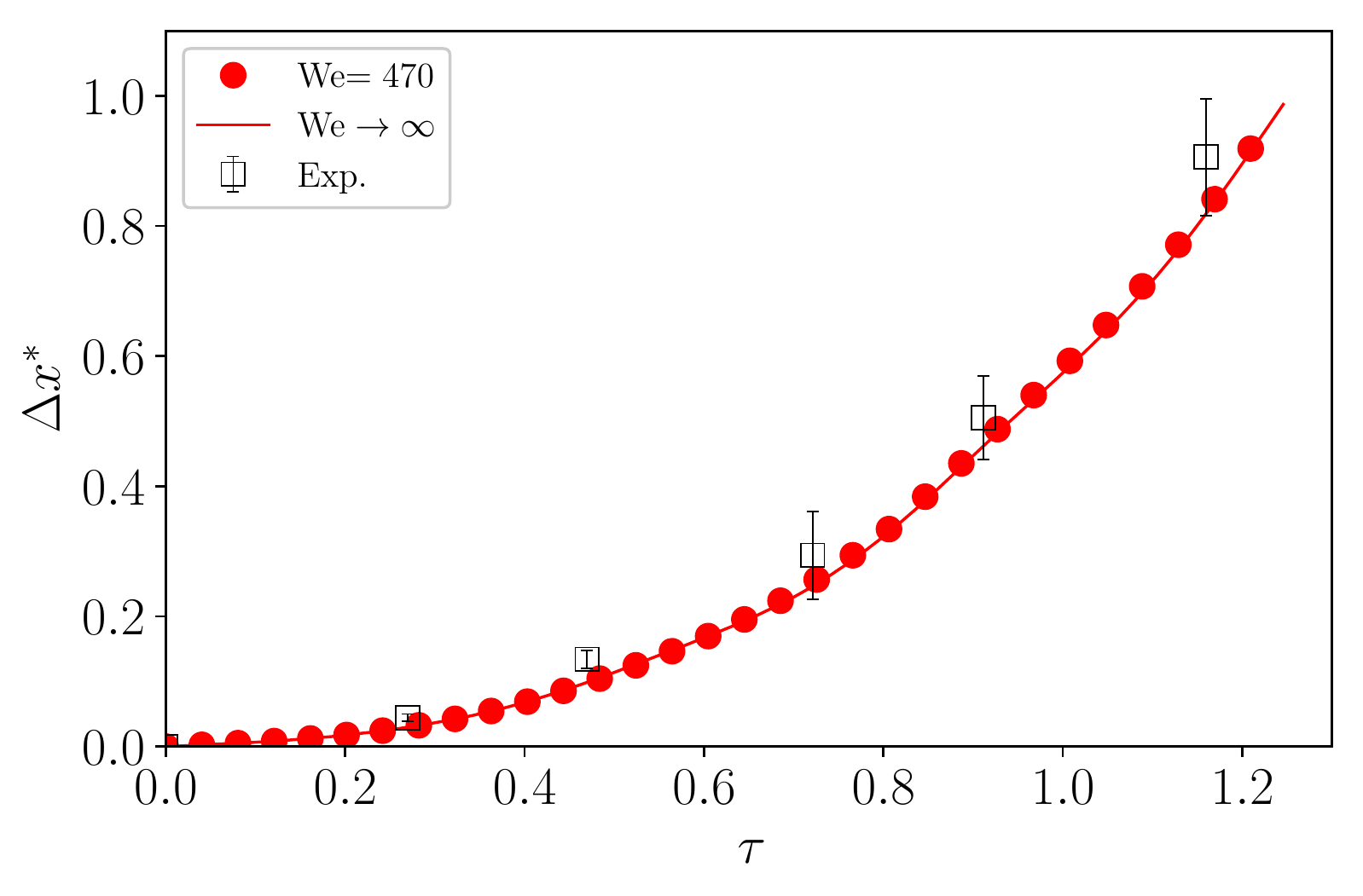}}
  \caption{Evolution of the center-of-mass from numerical results and experimental observations. Error bars display the sensitivity of the center-of-mass detection to the intensity threshold used in the binarization process. 
  }
  \label{fig:5}
\end{figure}

Further validation is provided in figure \ref{fig:5} where the droplet center-of-mass 
drift from the numerical simulation is compared with that measured in the experiments. 
From the numerical simulation the center-of-mass is computed 
as in \cite{meng2018numerical} using: 
\begin{equation}
\bm{x}_c=
\frac{\int_{\Omega_D} \alpha_l \rho_l \bm{x} {\rm{dV}}}{\int_{\Omega_D} \alpha_l \rho_l {\rm{dV}}}, 
\end{equation}
where $\Omega_D$ is the entire computational domain. 
On the experimental side, the evolution of the center-of-mass is computed 
from 2D-planar images,
due
to the line-integrated nature of images recorded by the shadowgraph.
The center-of-mass is determined by calculating the first order spatial
moment which is the intensity-weighted average of the pixel coordinates
constituting the droplet. This requires a binary image. The binarization is
performed by setting an intensity threshold to the image separating the droplet
from the the background.
Despite a slight deviation due to the 2D-planar
assumption and the threshold sensitivity, figure \ref{fig:5} shows a good
quantitative agreement between numerical results and the experiments. It can be
noticed that the surface tension has no discernible affect on the drift in the simulations.  This observation is consistent with the similar droplet morphology observed in both simulations at We=$470$ and $\bWe$.

\section{Formation of ligaments}
\label{sec:ligaments0}


For ${\rm{We}}>$10$^2$, the process of ligament formation is traditionally described
by the sheet-thinning mechanism proposed by \cite{liu1997analysis} on the basis of the work
of \cite{samuelsen1990experimental} for the breakup of a two-dimensional liquid sheet. Recently,
the formation of ligaments has been re-evaluated \citep{jalaal2014transient,
meng2018numerical} through 3D numerical simulations and an alternative driving mechanism
has been proposed namely, the transverse azimuthal modulation. To date, no
consensus has been found, and the ligament formation process has yet to be
understood.


\subsection{Ligament formation and azimuthal modulation}
\label{sec:ligaments}
\begin{figure}
  \centerline{\includegraphics[width=\textwidth]{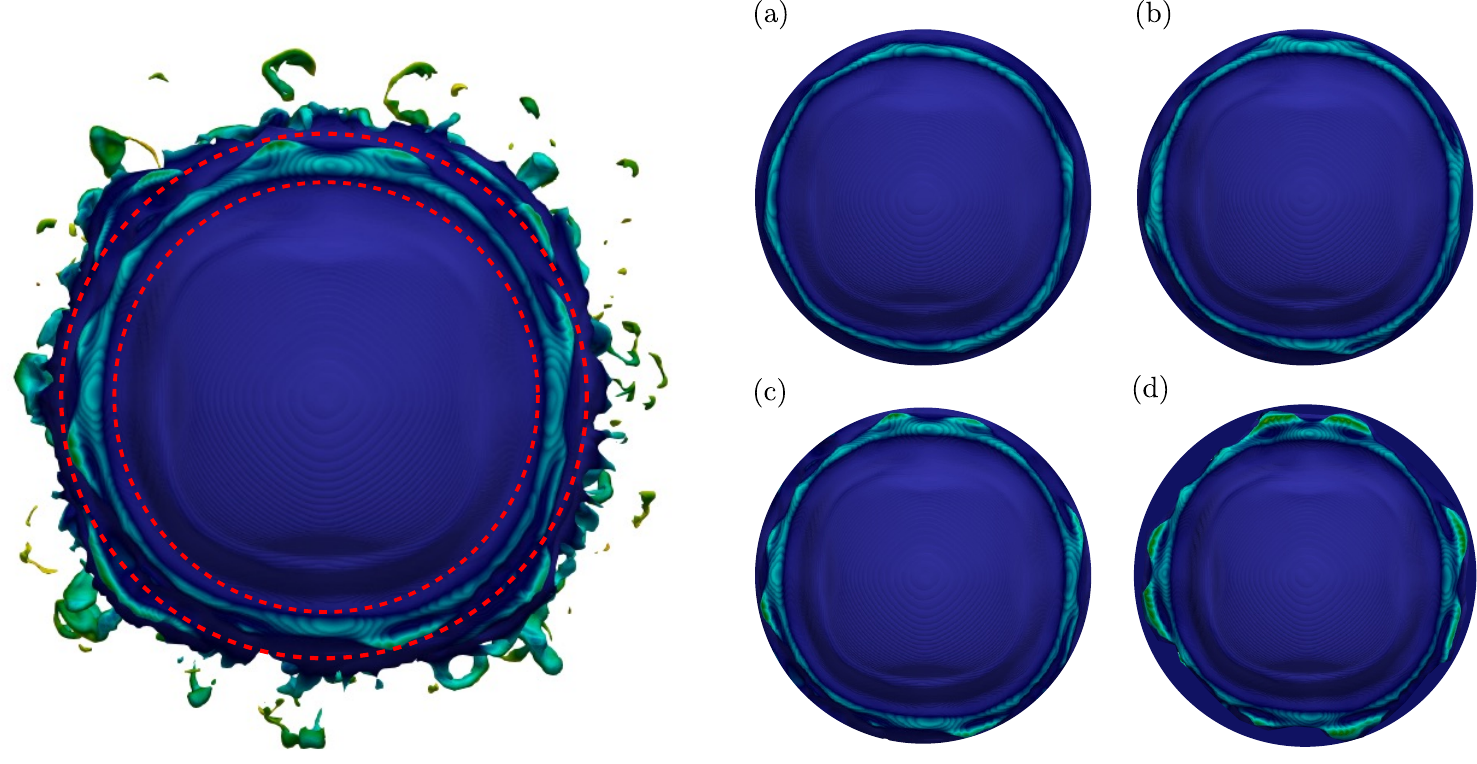}}
  \caption{Transverse azimuthal modulation for ${\rm{We}}$=470 (front view).
  Characteristic times are (a) 0.79, (b) 0.82, (c) 0.84 and (d) 0.86. Images (a)
  to (d) are cropped views.}
  \label{fig:crests}
\end{figure}
\begin{figure}
\centering
\begin{subfigure}[b]{0.3\textwidth}
\includegraphics[width=\textwidth]{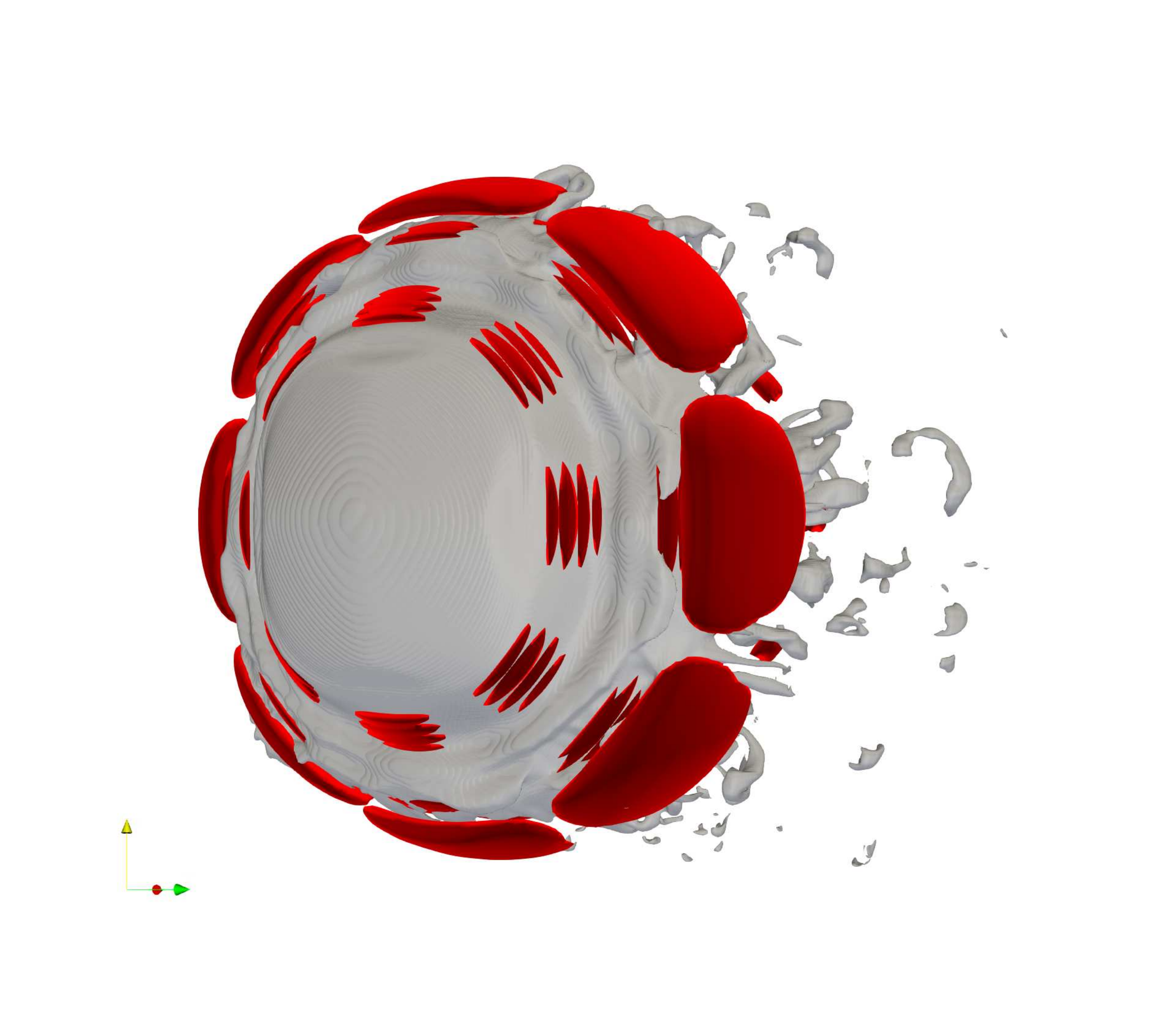}
\caption{m=4, $\tau\approx 0.85$}
\label{fig:st_fft_m40}
\end{subfigure}
\begin{subfigure}[b]{0.3\textwidth}
\includegraphics[width=\textwidth]{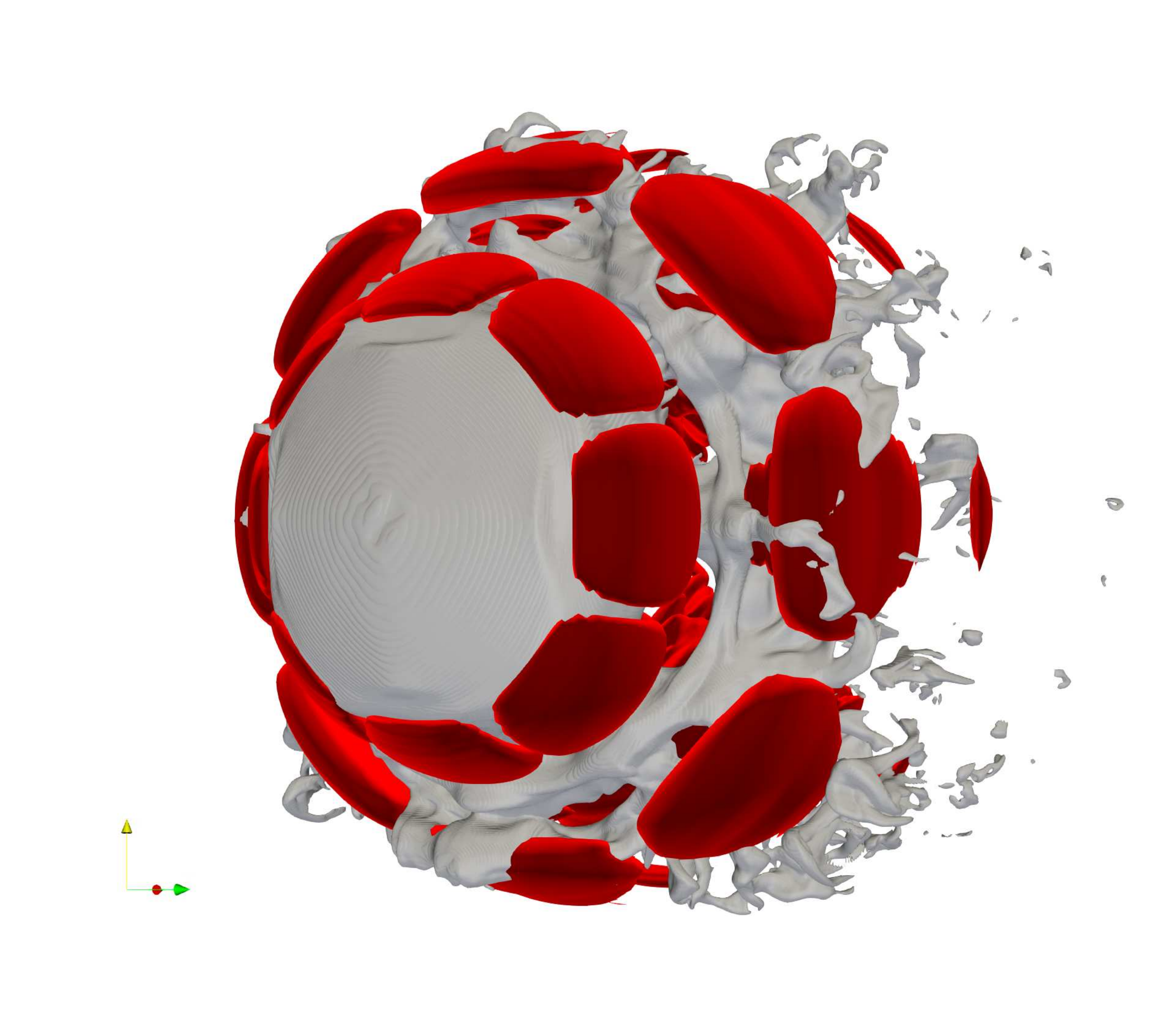}
\caption{m=4, $\tau\approx 0.97$}
\label{fig:st_fft_m41}
\end{subfigure}
\begin{subfigure}[b]{0.3\textwidth}
\includegraphics[width=\textwidth]{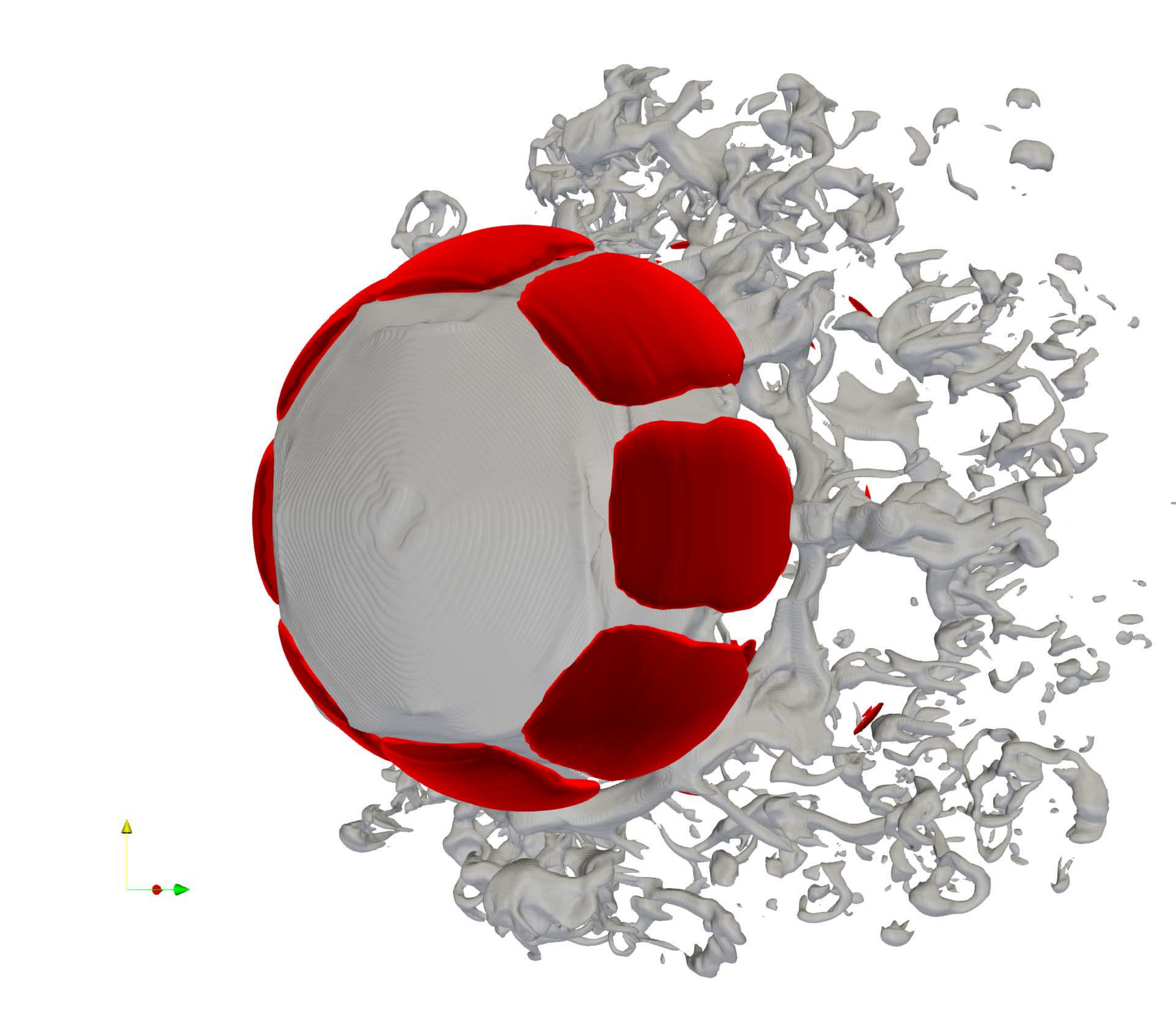}
\caption{m=4, $\tau\approx 1.05$}
\label{fig:st_fft_m42}
\end{subfigure}
\begin{subfigure}[b]{0.3\textwidth}
\includegraphics[width=\textwidth]{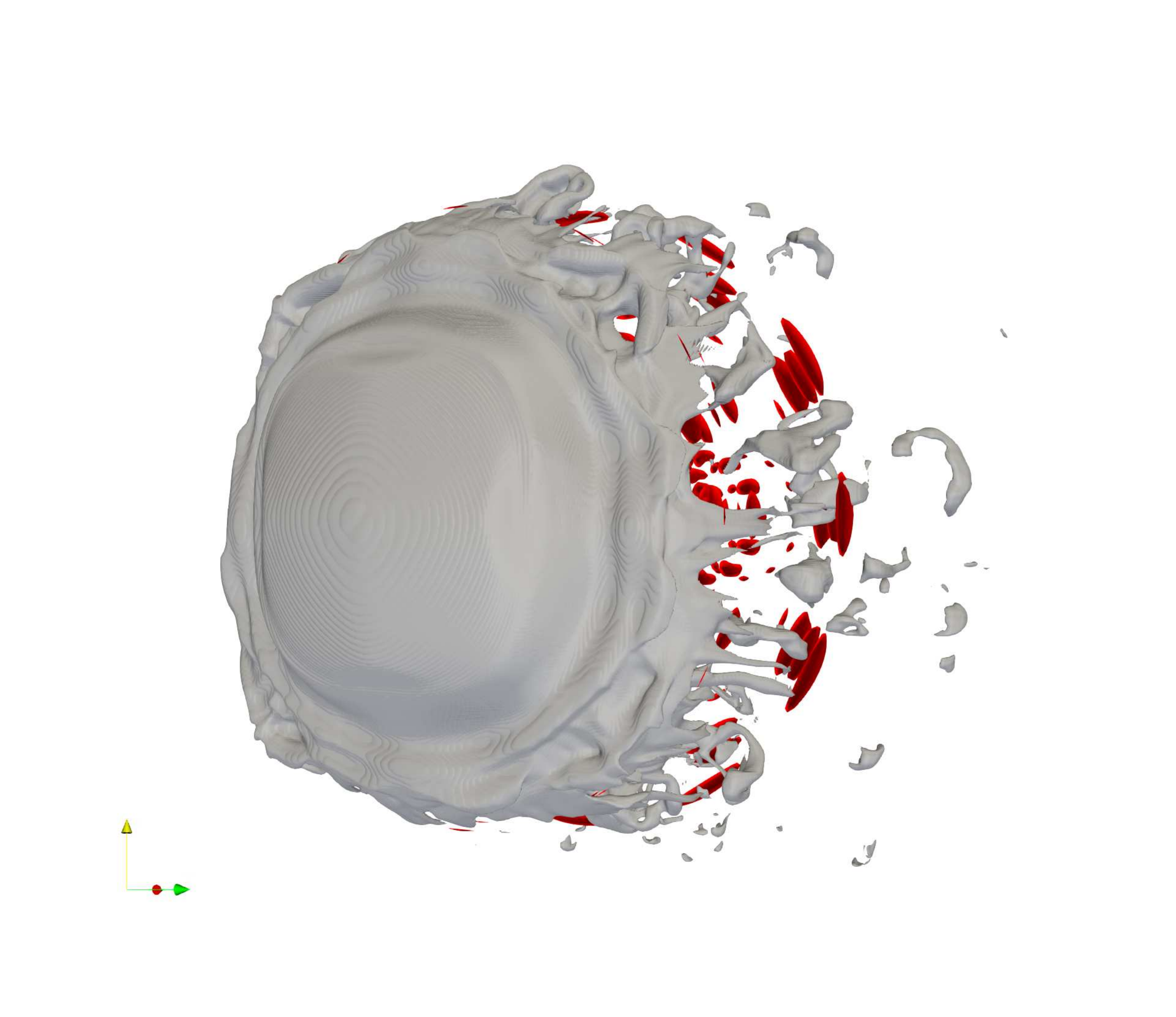}
\caption{m=6, $\tau\approx 0.85$}
\label{fig:st_fft_m60}
\end{subfigure}
\begin{subfigure}[b]{0.3\textwidth}
\includegraphics[width=\textwidth]{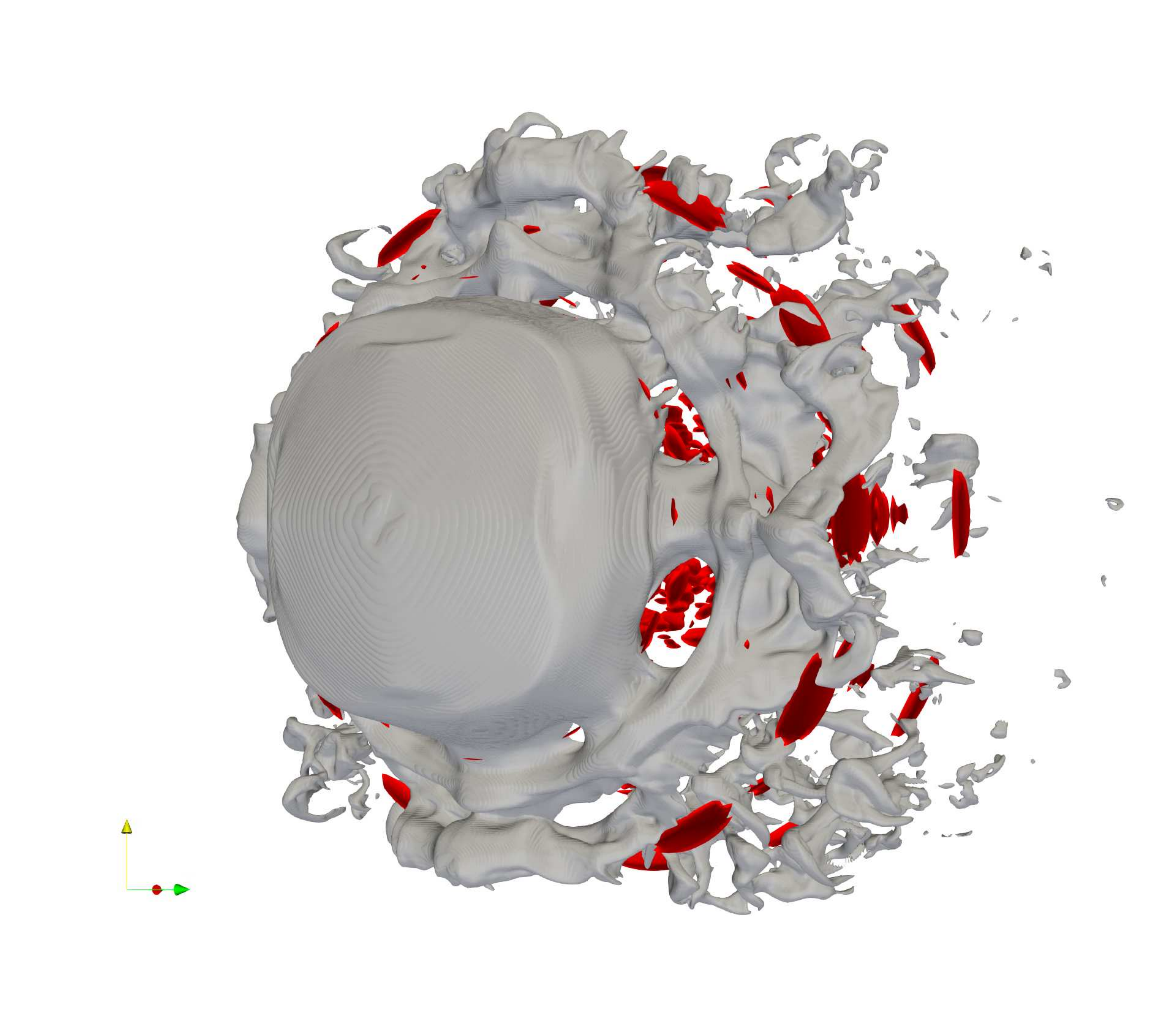}
\caption{m=6, $\tau\approx 0.97$}
\label{fig:st_fft_m61}
\end{subfigure}
\begin{subfigure}[b]{0.3\textwidth}
\includegraphics[width=\textwidth]{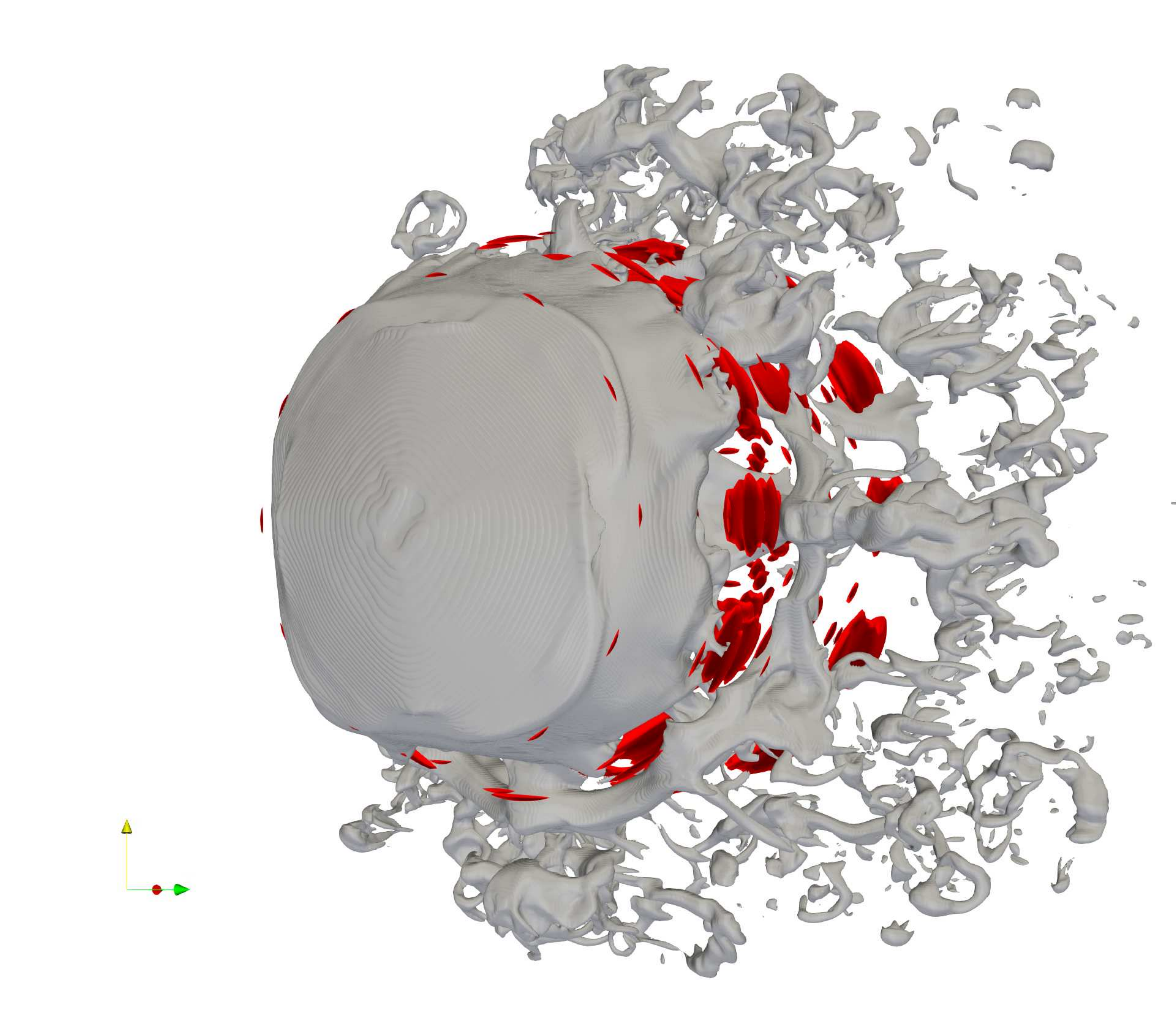}
\caption{m=6, $\tau\approx 1.05$}
\label{fig:st_fft_m62}
\end{subfigure}
\begin{subfigure}[b]{0.3\textwidth}
\includegraphics[width=\textwidth]{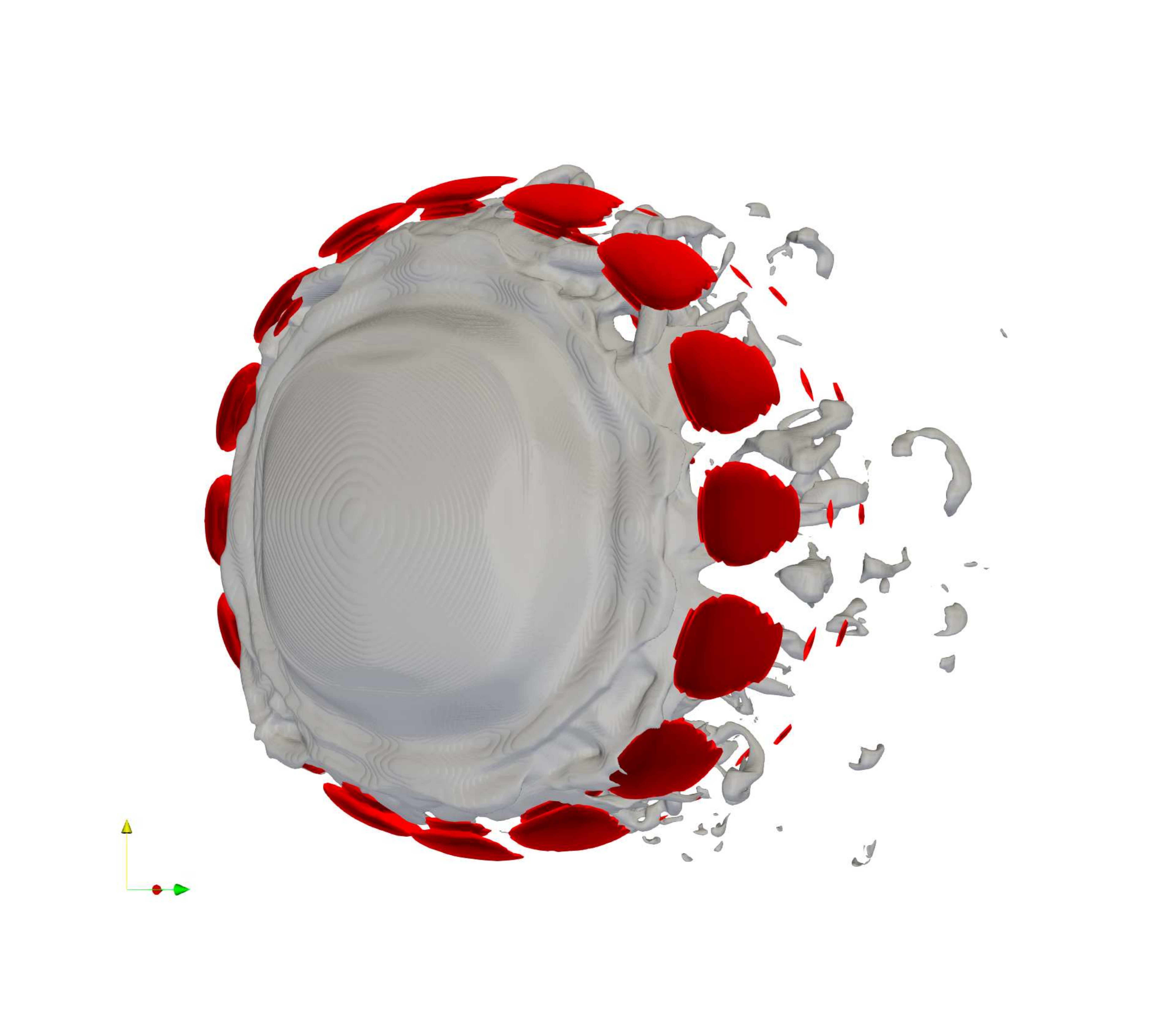}
\caption{m=8, $\tau\approx 0.85$}
\label{fig:st_fft_m80}
\end{subfigure}
\begin{subfigure}[b]{0.3\textwidth}
\includegraphics[width=\textwidth]{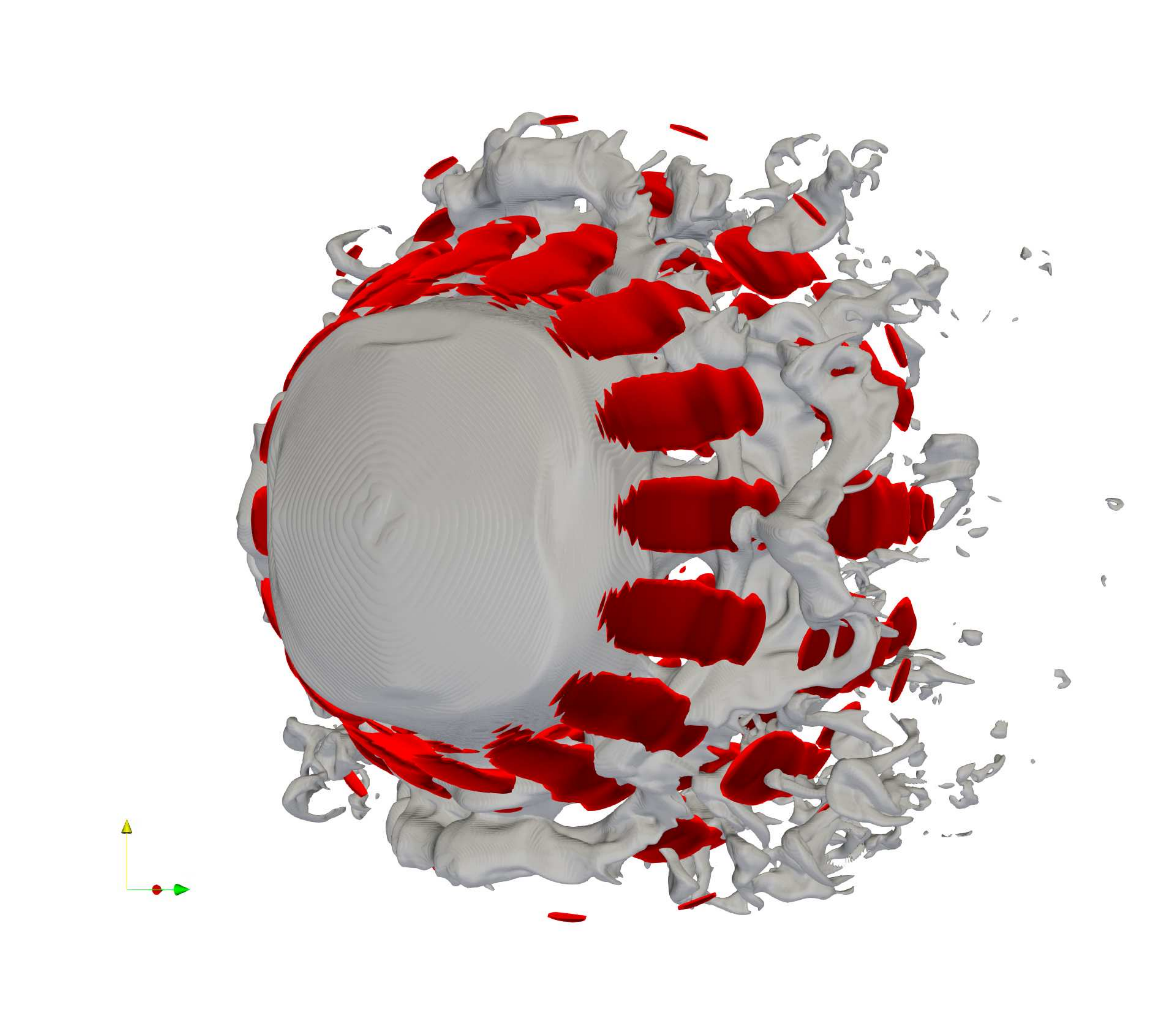}
\caption{m=8, $\tau\approx 0.97$}
\label{fig:st_fft_m81}
\end{subfigure}
\begin{subfigure}[b]{0.3\textwidth}
\includegraphics[width=\textwidth]{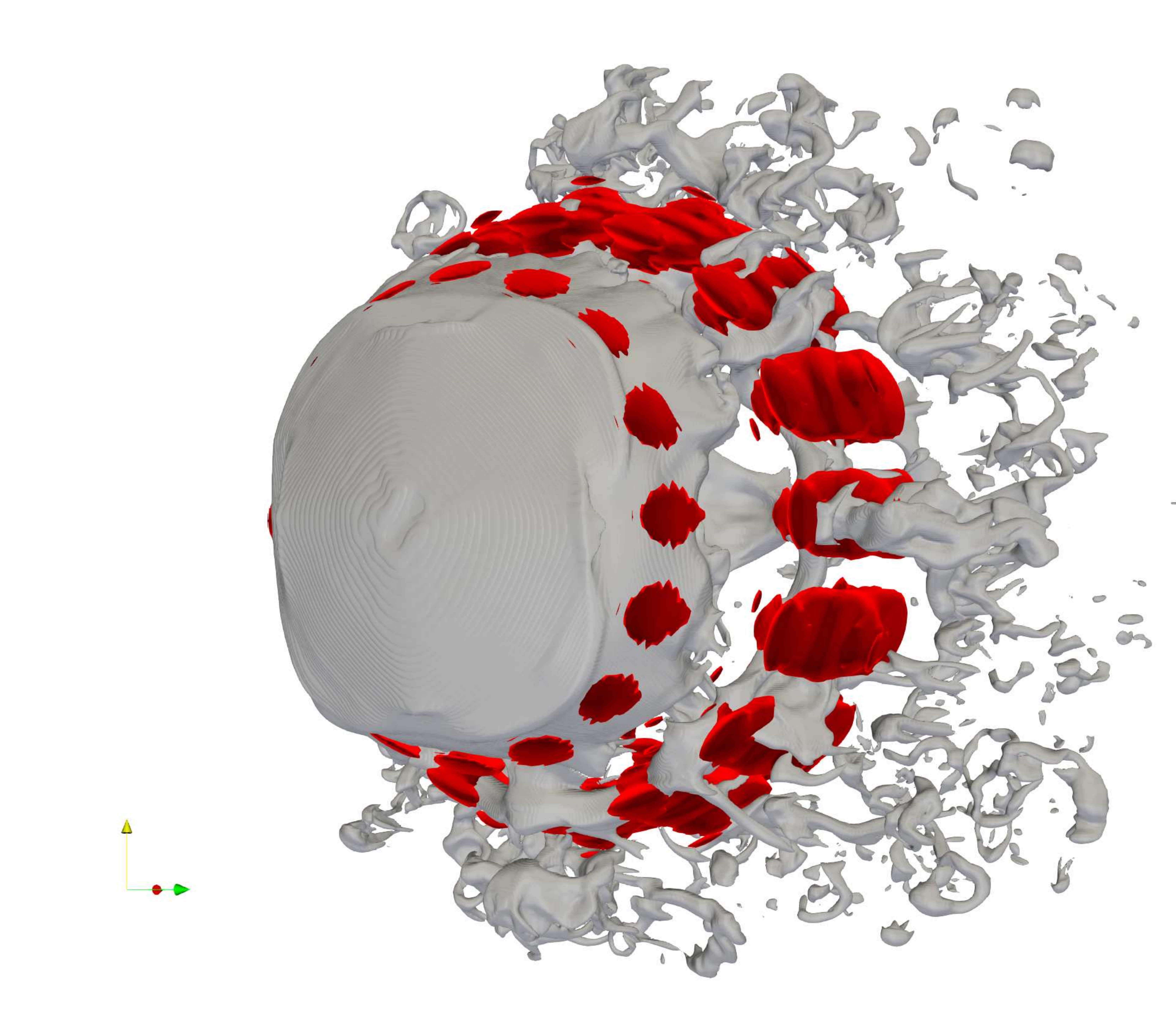}
\caption{m=8, $\tau\approx 1.05$}
\label{fig:st_fft_m82}
\end{subfigure}
\caption{Isosurfaces of the $m$-th azimuthal mode $\kappa_{m}$ (red)  and
isosurfaces of the volume fraction $\alpha_l$=0.01 (grey) for ${\rm{We}}$=470.}
\label{fig:FFT_st}
\end{figure}
\begin{figure}
\begin{subfigure}[b]{0.3\textwidth}
\includegraphics[width=\textwidth]{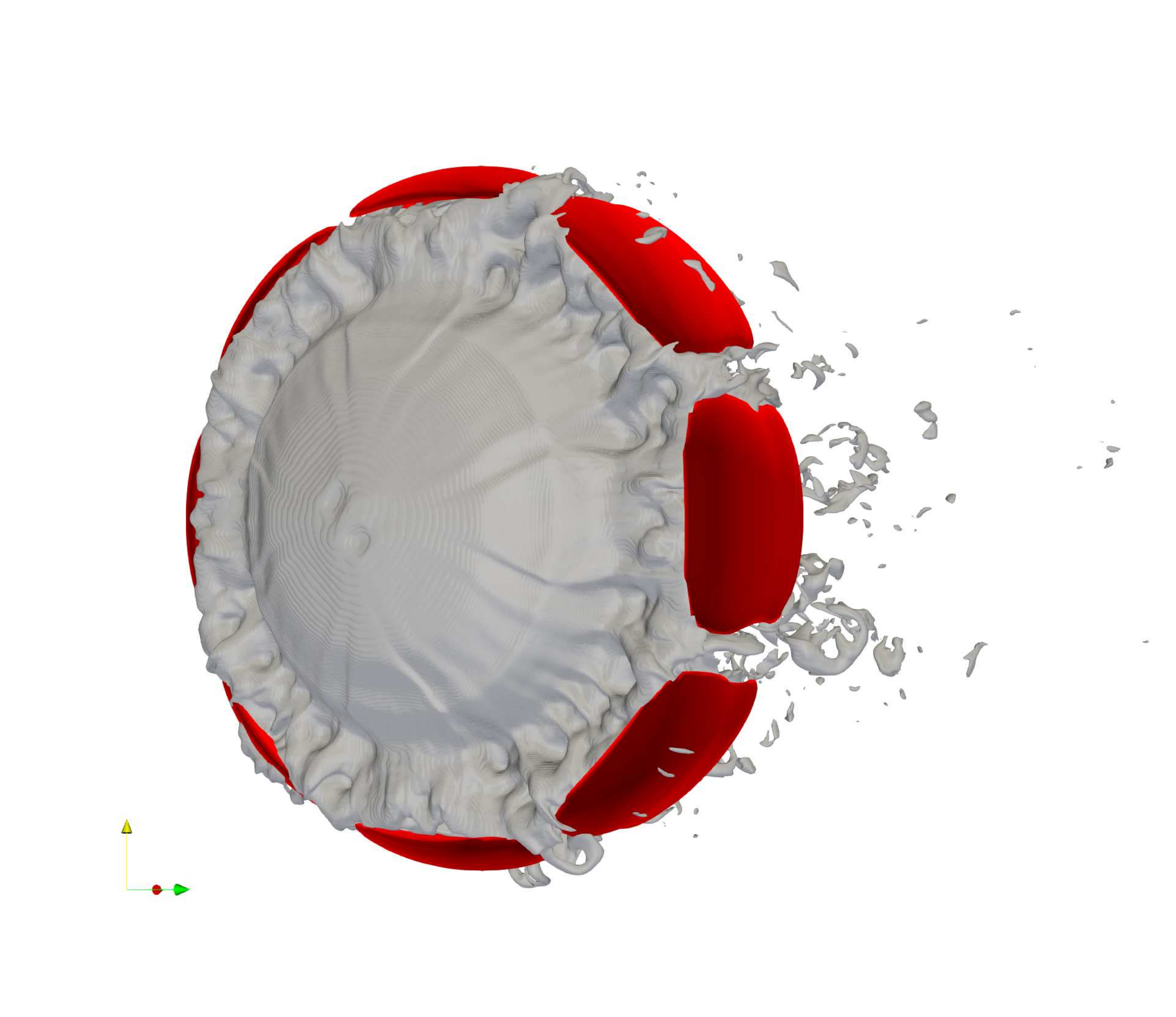}
\caption{m=4, $\tau\approx 0.85$}
\label{fig:nst_fft_m40}
\end{subfigure}
\begin{subfigure}[b]{0.3\textwidth}
\includegraphics[width=\textwidth]{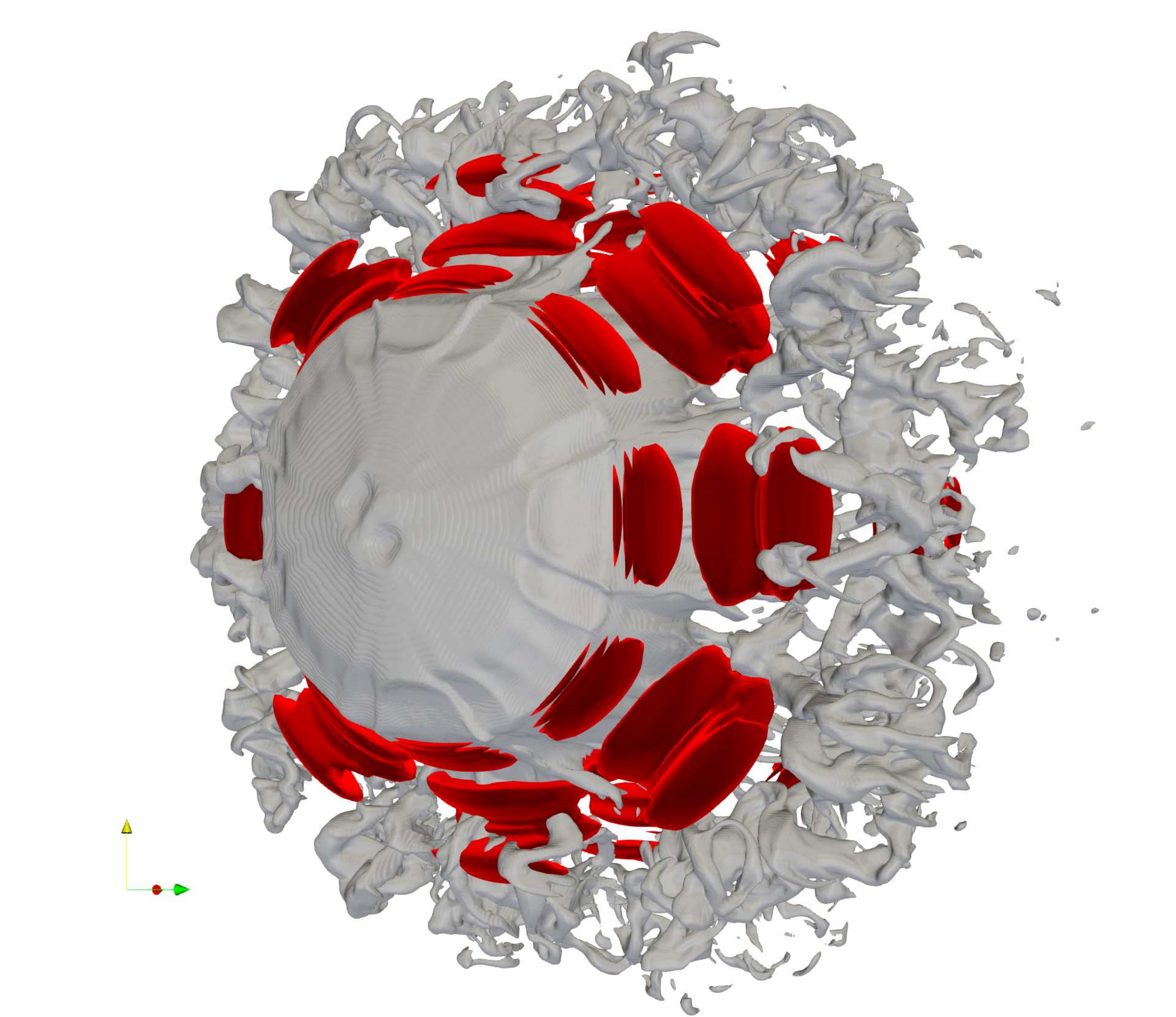}
\caption{m=4, $\tau\approx 0.97$}
\label{fig:nst_fft_m41}
\end{subfigure}
\begin{subfigure}[b]{0.3\textwidth}
\includegraphics[width=\textwidth]{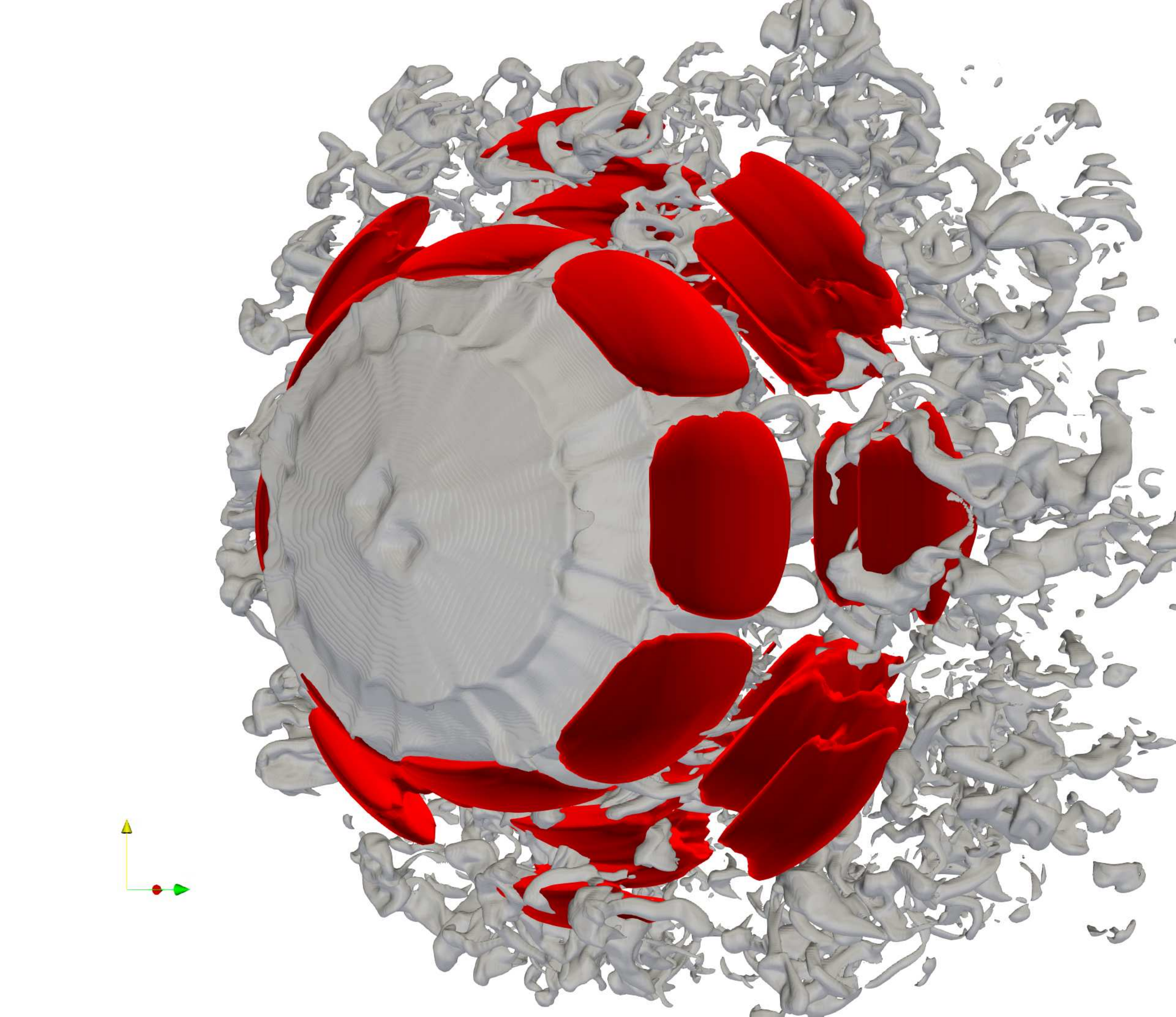}
\caption{m=4, $\tau\approx 1.05$}
\label{fig:nst_fft_m42}
\end{subfigure}
\\
\begin{subfigure}[b]{0.3\textwidth}
\includegraphics[width=\textwidth]{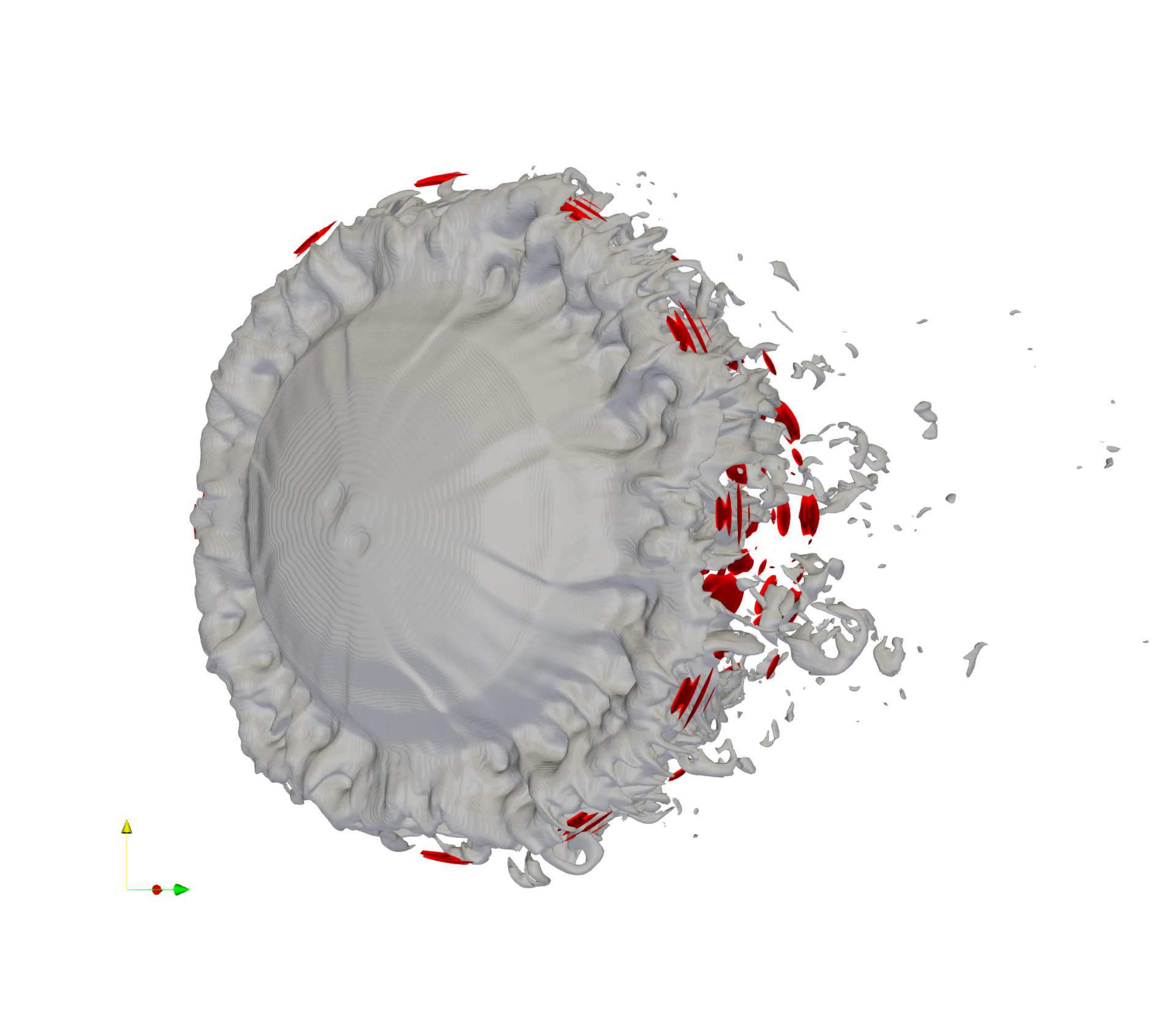}
\caption{m=6, $\tau\approx 0.85$}
\label{fig:nst_fft_m60}
\end{subfigure}
\begin{subfigure}[b]{0.3\textwidth}
\includegraphics[width=\textwidth]{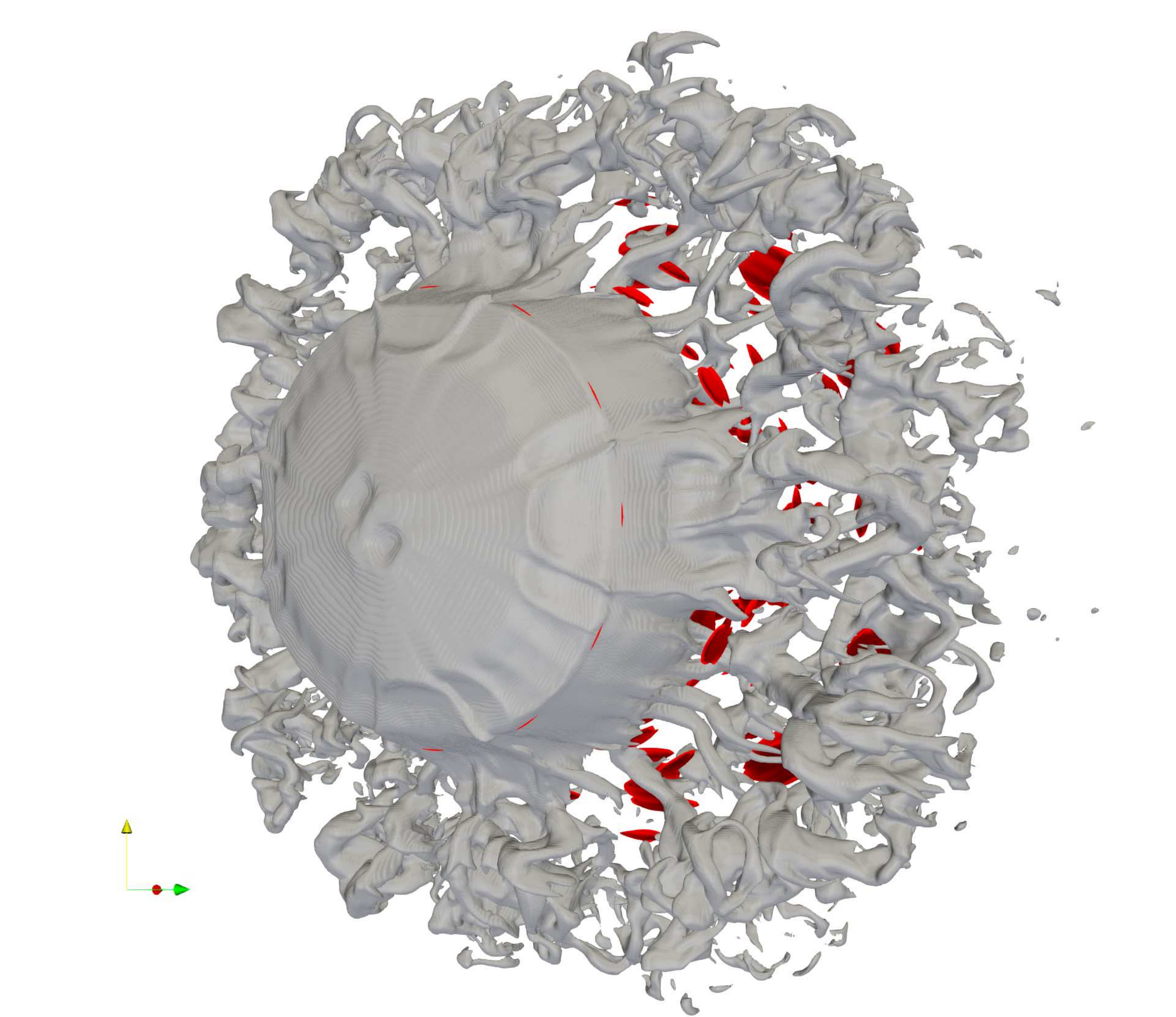}
\caption{m=6, $\tau\approx 0.97$}
\label{fig:nst_fft_m61}
\end{subfigure}
\begin{subfigure}[b]{0.3\textwidth}
\includegraphics[width=\textwidth]{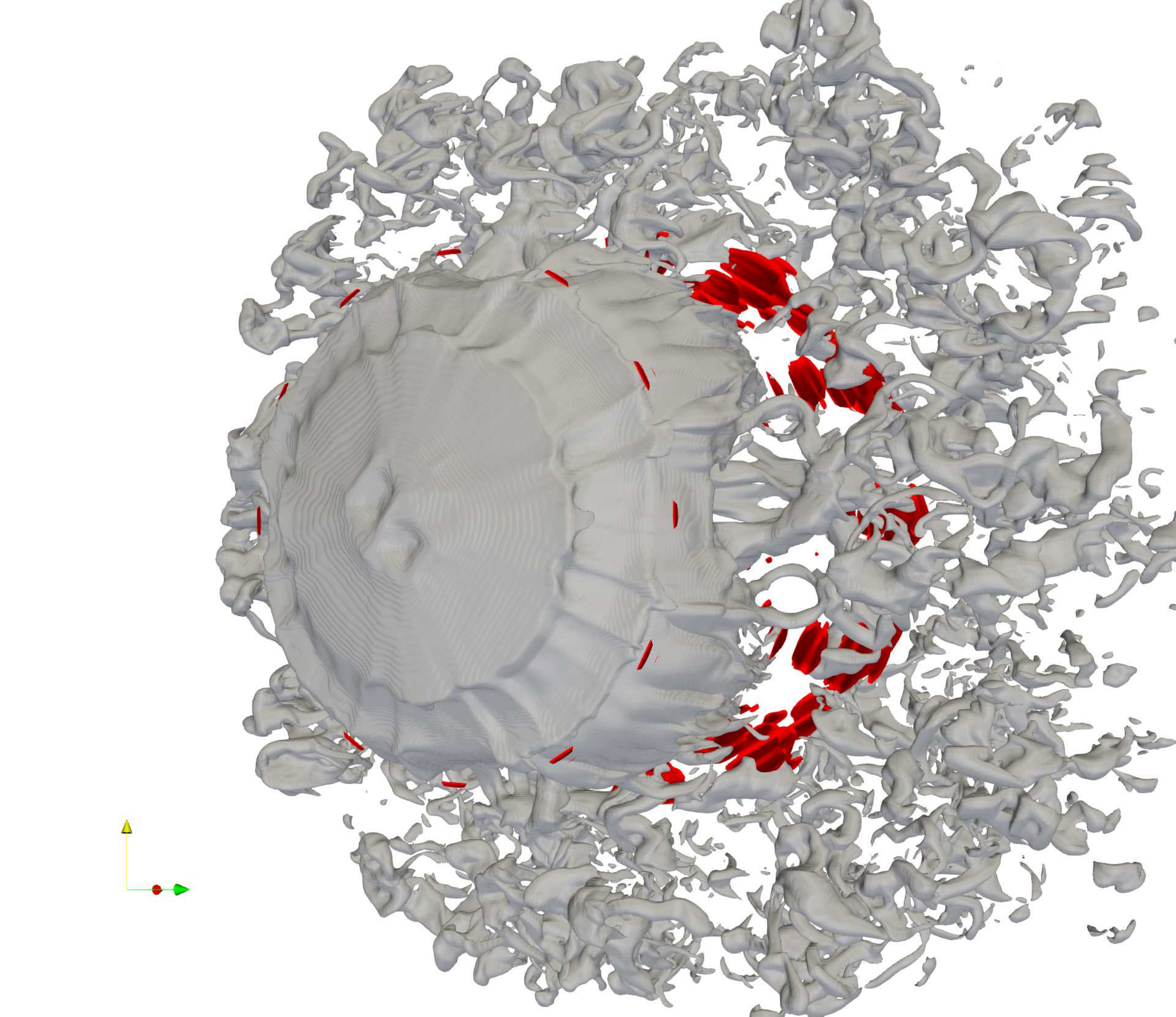}
\caption{m=6, $\tau\approx 1.05$}
\label{fig:nst_fft_m62}
\end{subfigure}
\\
\begin{subfigure}[b]{0.3\textwidth}
\includegraphics[width=\textwidth]{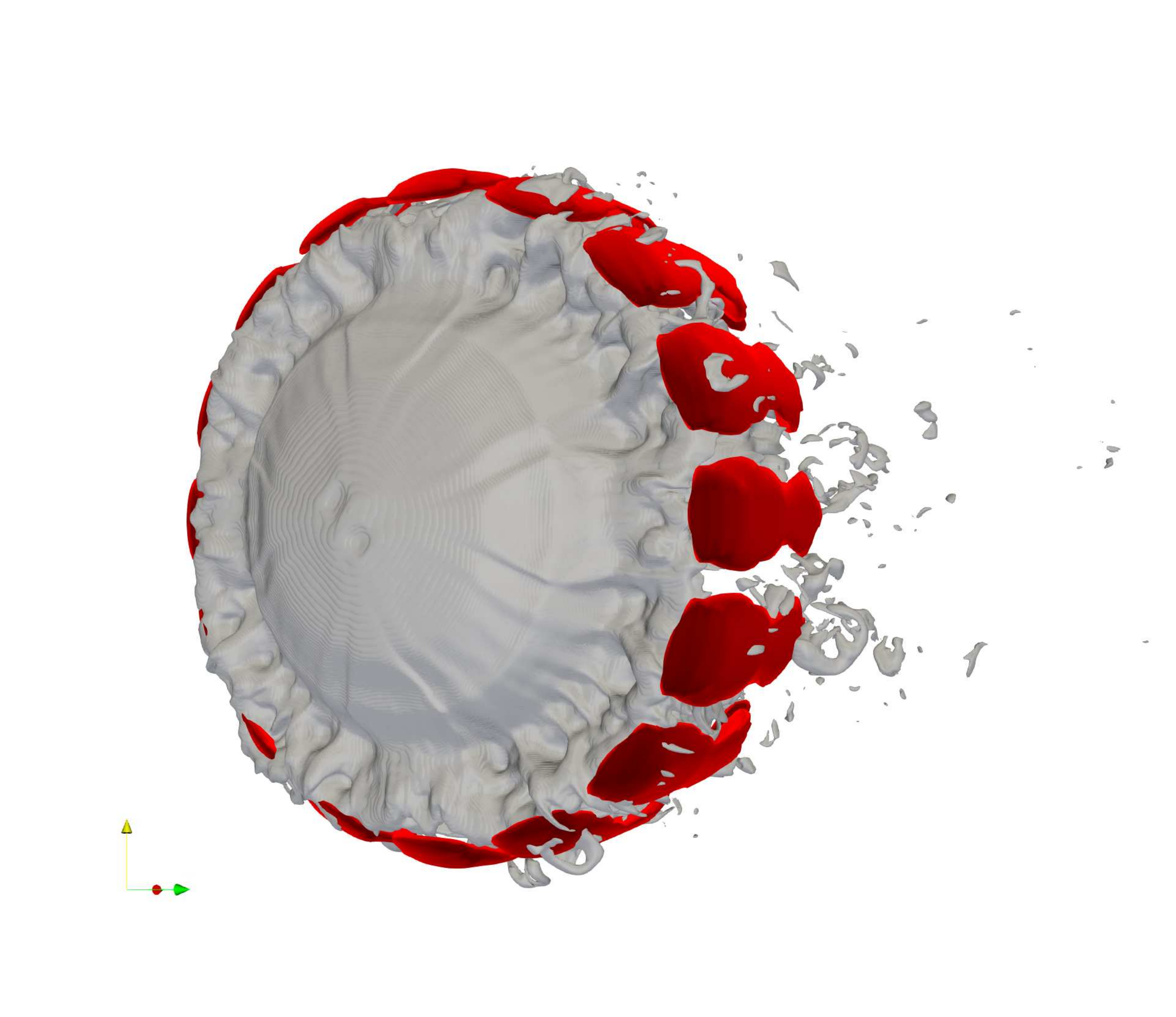}
\caption{m=8, $\tau\approx 0.85$}
\label{fig:nst_fft_m80}
\end{subfigure}
\begin{subfigure}[b]{0.3\textwidth}
\includegraphics[width=\textwidth]{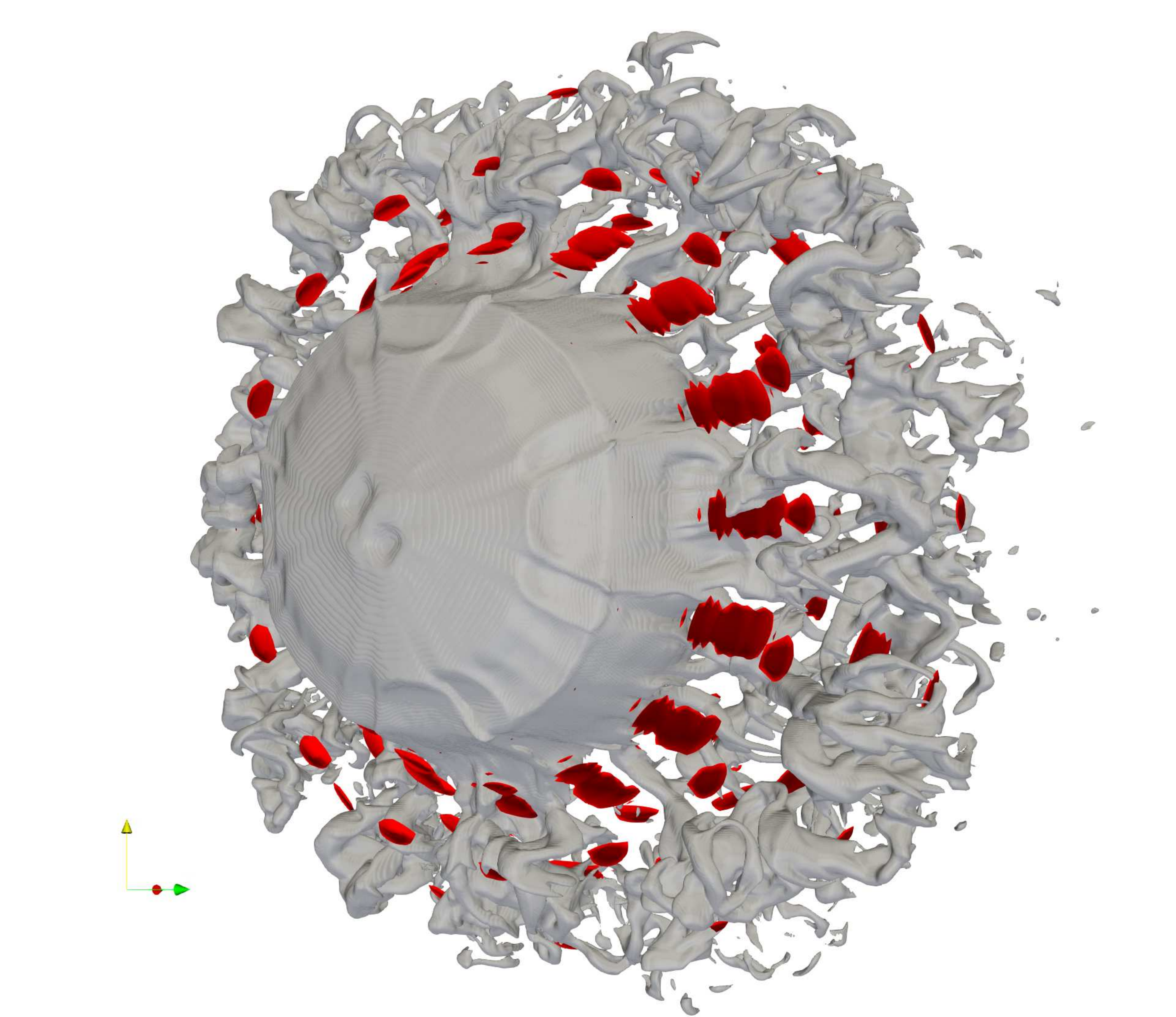}
\caption{m=8, $\tau\approx 0.97$}
\label{fig:nst_fft_m81}
\end{subfigure}
\begin{subfigure}[b]{0.3\textwidth}
\includegraphics[width=\textwidth]{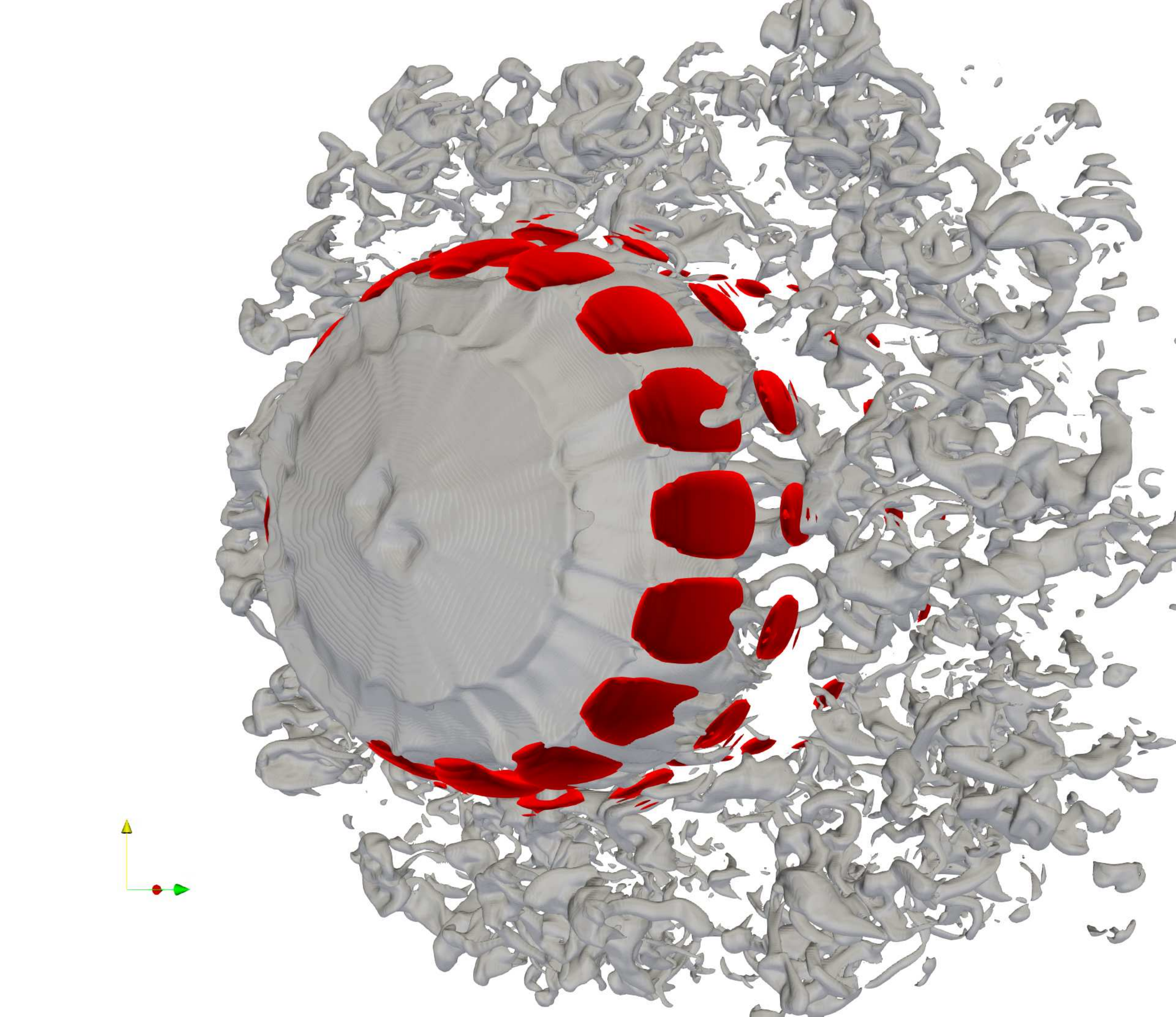}
\caption{m=8, $\tau\approx 1.05$}
\label{fig:nst_fft_m81}
\end{subfigure}
\caption{Isosurfaces of the $m$-th azimuthal mode $\kappa_{m}$ (red) and
isosurfaces of the volume fraction $\alpha_l$=0.01 (grey) for $\bWe$.}
\label{fig:FFT_Nst}
\end{figure}
Exposing the water droplet to a high-speed gas flow can 
induce a large difference in the velocities between 
the two fluids and thus destabilize their interface. 
Depending on the flow conditions, the interface may be more susceptible 
to Rayleigh-Taylor or Kelvin-Helmholtz instabilities. 
The canonical Rayleigh-Taylor instability (RTI)
occurs when a heavy fluid is suspended above a lighter fluid and both are
subjected to acceleration. The interface begins to oscillate
with alternating acceleration directions and is unstable when the acceleration
is oriented towards the heavier fluid \citep{RTI}. 
On the other hand, Kelvin-Helmholtz
instability (KHI) arises due to either shear in a single fluid system or due to 
a velocity difference across the interface of two fluids, which results in
propagating waves, and typically rollup into vortices, along the interface \citep{chandrasekharhydrodynamic}.

For low Weber numbers (${\rm{We}} \sim$ 20), it is thought that the RTI is
responsible for the bag-shape structure, which is attached to a thicker
toroidal rim.  Increasing the Weber number up to $80$, the standard bag
morphology evolves to more complex bag-structures, namely the bag-and-stamen
and multi-bag modes, still believed to be driven by RTI
\citep{jain2015secondary}. Ultimately, bags undergo a piercing mechanism
\citep{guildenbecher2009secondary}. 

For higher Weber numbers there is less of a consensus but the markedly different 
droplet morphologies are also attributed to mechanisms other than RTI piercing.
In the work of \citet{jain2015secondary} on the secondary atomization, the
authors numerically investigate the shear-stripping regime, i.e., sheet-thinning
regime, by running simulations at ${\rm{We}}$=120. Arguing that high inertial forces
overcome the restoring effect of the surface tension, the authors state that the
development of a potential bag on the rim is hampered by the stripping process
occurring at the droplet equatorial plan. This indirectly exonerates the
Rayleigh-Taylor piercing mechanism in the ligament formation. The authors
suggest that the ligament formation owes to the high-speed gas flow over the
droplet periphery, which results in transverse RTI. The crests of the
transverse instability are deformed into ligaments by being stretched with the
flow as described by \cite{marmottant2004spray}.
Following the work of \citet{marmottant2004spray} on the atomization of a
liquid jet in coaxial flow, the recent work of \citet{jalaal2014transient}
suggests that axisymmetric waves, propagating on the droplet surface, are the result
of KHI and that their acceleration leads to a transverse azimuthal modulation, 
which can be viewed as RTI.
As a result of such KHI-RTI combination,
streamwise ligaments are formed and subsequently fragmented into droplets.
In addition, their 3D numerical simulations 
for Weber numbers up to 200 show a good qualitative agreement with their 
conjecture of an azimuthal modulation due to RTI. In a attempt to provide quantitative
evidence, they compared the most-amplified wavenumbers, deduced from their
simulations, with theoretical predictions but failed to find any conclusive
quantitative agreement. 
In another attempt to find quantitative evidence of azimuthal RTI, 
\citet{meng2018numerical} performed a 3D simulation for $\bWe$.
In line with \cite{jalaal2014transient}, these simulations show a loss of axisymmetry, 
but a Fourier analysis of the velocity field reveals broadband instability growth 
for all modes and hence does not provide further evidence of transverse RTI.

In the present numerical simulations for ${\rm{We}}$ = 470, we also observe a loss of
axisymmetry of the liquid sheet propagating on the droplet rim before transverse
corrugations arise at the interface.
In particular, as apparent from figure \ref{fig:crests}, we can see 
a non-uniform growth rate of the transverse corrugations, 
which results in variable crest amplitudes, indicating a transverse azimuthal modulation of the droplet. 
However, concerning the relation of the transverse instability with the
ligament formation, a mismatch between the wavelength and the number of
crests and ligaments does not seem to confirm the mechanism proposed by
\citet{jalaal2014transient}. According to their conjecture,
ligaments arise from the
stretching of the interface corrugation crests due to aerodynamic forces. This
implies that the number of ligaments is equal to the number of crests. As
shown in figure \ref{fig:crests}, the number of crests observed in our simulation
does not directly correspond to the eight ligaments, which are ultimately forming
in the course of the breakup.

\begin{figure}
  \centerline{\includegraphics[width=0.5\textwidth]{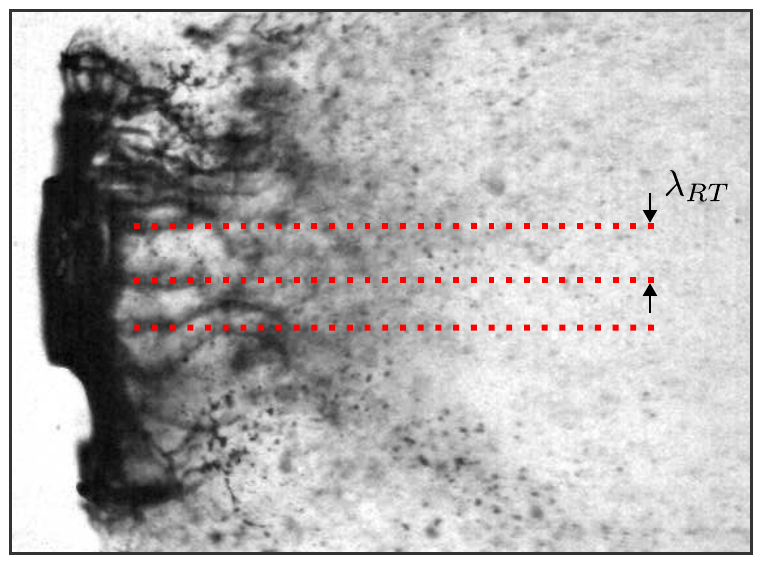}}
  \caption{Measurement of ligament wavenumber.}
  \label{fig:wavenumber}
\end{figure}
For further investigation of the observed azimuthal modulation and to identify 
the modes which lead to the formation of the ligaments as well as to determine the effect 
of surface tension on these modes,
we decompose the flow field around the droplet into azimuthal Fourier modes.
To that end, the Cartesian mesh is interpolated onto a cylindrical mesh, where 
the resolutions in
azimuthal direction $\theta$, radial direction $r$ and streamwise direction $x$
are kept similar to the Cartesian mesh.
Subsequently, the azimuthal Fourier coefficients for each mode $\hat{\bm{u}}(x,r,t)$ are obtained 
by Fourier transforms in $\theta$ direction. We use an energy metric of the velocity for each 
mode, which is given by 
\begin{equation}
    \hat{\kappa}_m=|\hat{u}_{x,m}|^2+|\hat{u}_{r,m}|^2+|\hat{u}_{\theta,m}|^2.
\end{equation}
Transforming $\hat{\kappa}_m$ back into physical space yields the azimuthal Fourier
mode $\kappa_m$. Isosurfaces of selected azimuthal modes are superimposed with the isosurfaces
of the volume fraction in figure \ref{fig:FFT_st}, 
which shows that the azimuthal wavenumber $m=4$ and its harmonic $m=8$
are most pronounced on the droplet surface. 
Modes corresponding to wavenumbers other than $m=4$ or $m=8$, 
develop in the wake on the back of the droplet and do not appear to be 
responsible for the azimuthal modulation and deformation of the droplet itself. 
%
%
%
This is exemplarily shown for $m=6$  in figures \ref{fig:st_fft_m60}-\ref{fig:st_fft_m62} but similarly
observed for all other wavenumbers but $m=4$ and $m=8$.
Strikingly, the wavenumbers $m=4$ and its harmonic $m=8$ do correspond 
to the wavenumber of the ligaments and thus directly relate the azimuthal modulation  
of the flow field to the formation of the ligaments.

Experimentally, we measure the mean wavenumber of the ligaments by manual post-processing of the 
shadowgraphs and measuring the distance between ligaments, where a ligament is defined as a coherent column of liquid as exemplified in figure \ref{fig:wavenumber}. 
This procedure eventually yields a mean wavenumber 
of the ligaments of $ m_{{\rm{exp}}}\sim 8.5$
, which is in excellent 
agreement with what is observed in our simulations. This further establishes 
validity of both our experiments and the numerical simulations.

\begin{figure}
  \centerline{\includegraphics[width=\textwidth]{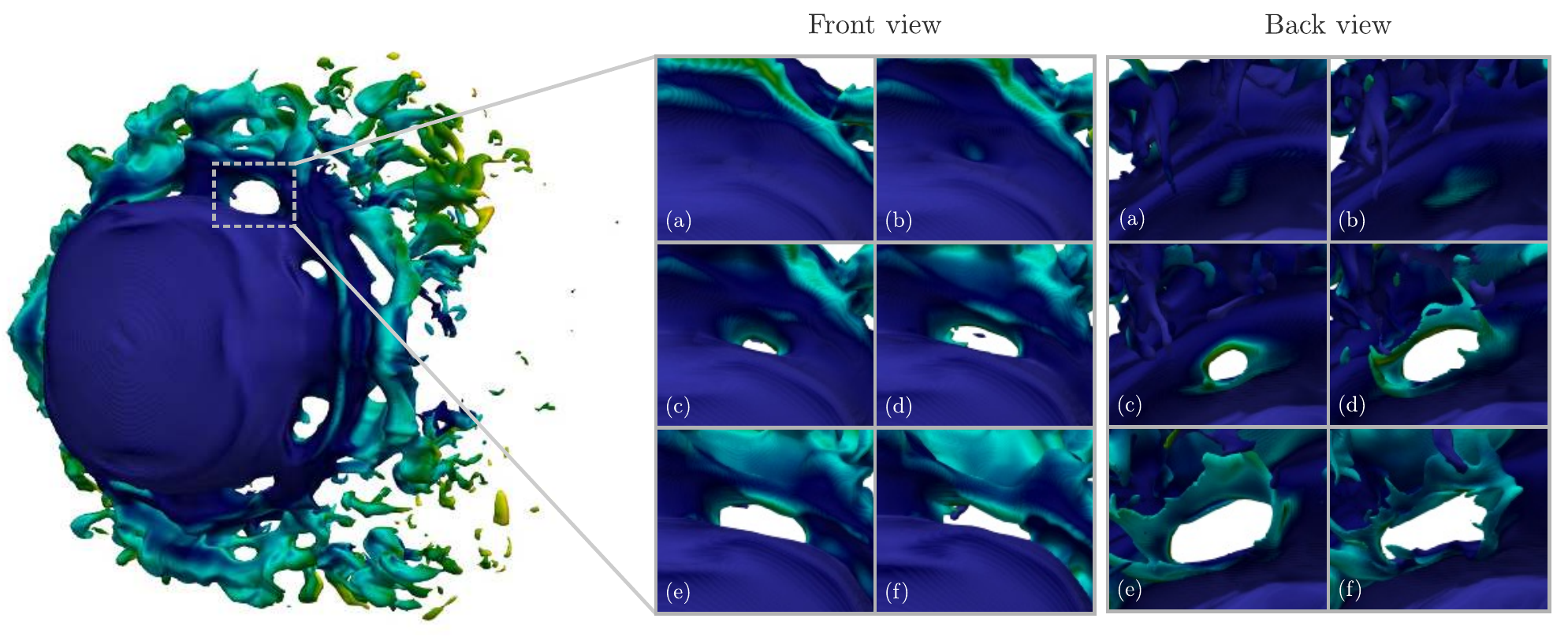}}
  \caption{Piercing mechanisms for ${\rm{We}}$=470 (front and back views). Characteristic times are (a) 0.88, (b) 0.90, (c) 0.92, (d) 0.94, (e) 0.96 and (f) 0.98.}
  \label{fig:RTP}
\end{figure}

When studying the evolution of the azimuthal structures of $m=4$ and $m=8$ in
our numerical simulations, one can observe that these structures are formed at
relatively early stages of the breakup and develop on the droplet surface.
Subsequently, the azimuthal modes appear to deform the droplet surface, which leads
to azimuthally distributed bag-like structures with wavenumber $m=8$.
At later stages, these bag-like structures develop further and grow 
as they are subjected to the high-speed gas flow surrounding the droplet.
Eventually, the aerodynamic forces are able to overcome the restoring effect
of surface tension and pierce or rupture the bag-like structures, yielding 
ligaments, which, in first instance, remain attached to the circular rim before 
they detach at later stages.
This piercing mechanism is illustrated by the detailed snapshots in figure \ref{fig:RTP}, 
which show the evolution and rupture of the bag-like structures.
It is also evident that eventually these ruptured pockets form the ligaments.

It is interesting to compare this picture to the We$_{\sigma=0}$ case, which is shown in figure \ref{fig:FFT_Nst}.  We can again observe that, as 
in the case of a finite Weber number, the only
modes that are acting on the droplet surface correspond to the wavenumbers
$m=4$  and $m=8$. This suggests that the cause and origin of these structure are
independent of capillary effects. However, there are obvious differences in the subsequent evolution.  
In particular, it is apparent that the azimuthal disturbances are less amplified, and less able to 
modulate 
the cylindrical liquid curtain that is formed around the droplet core during the course of the breakup.
Hence, in contrast to the finite Weber number case, there are no prominent
bag-like structures, and the liquid curtain remains unruptured and intact,
before it eventually disintegrates into
fine-scale structures. 
This observation is also in good qualitative agreement with what was observed by others \citep{marmottant2004spray,jalaal2014transient}
%
who conjecture that the transverse destabilisation of the liquid rim owes to the
Rayleigh-Taylor instability, which leads to an infinitesimally small wavelength
for negligible surface-tension values ($\sigma \to$ 0) as the RTI wavelength is
proportional to the root of the surface tension. 

While, we have not been able to infer a direct relation to RTI from our simulations, 
our observations do suggest that transverse azimuthal modulation plays a crucial role in the formation and subsequent shedding of ligaments.

\subsection{Recurrent breakup and vortex shedding}
\begin{figure}
\centering
\begin{subfigure}[t]{0.45\textwidth}
\includegraphics[clip, trim=20cm 2cm 5cm 2cm,width=\textwidth]{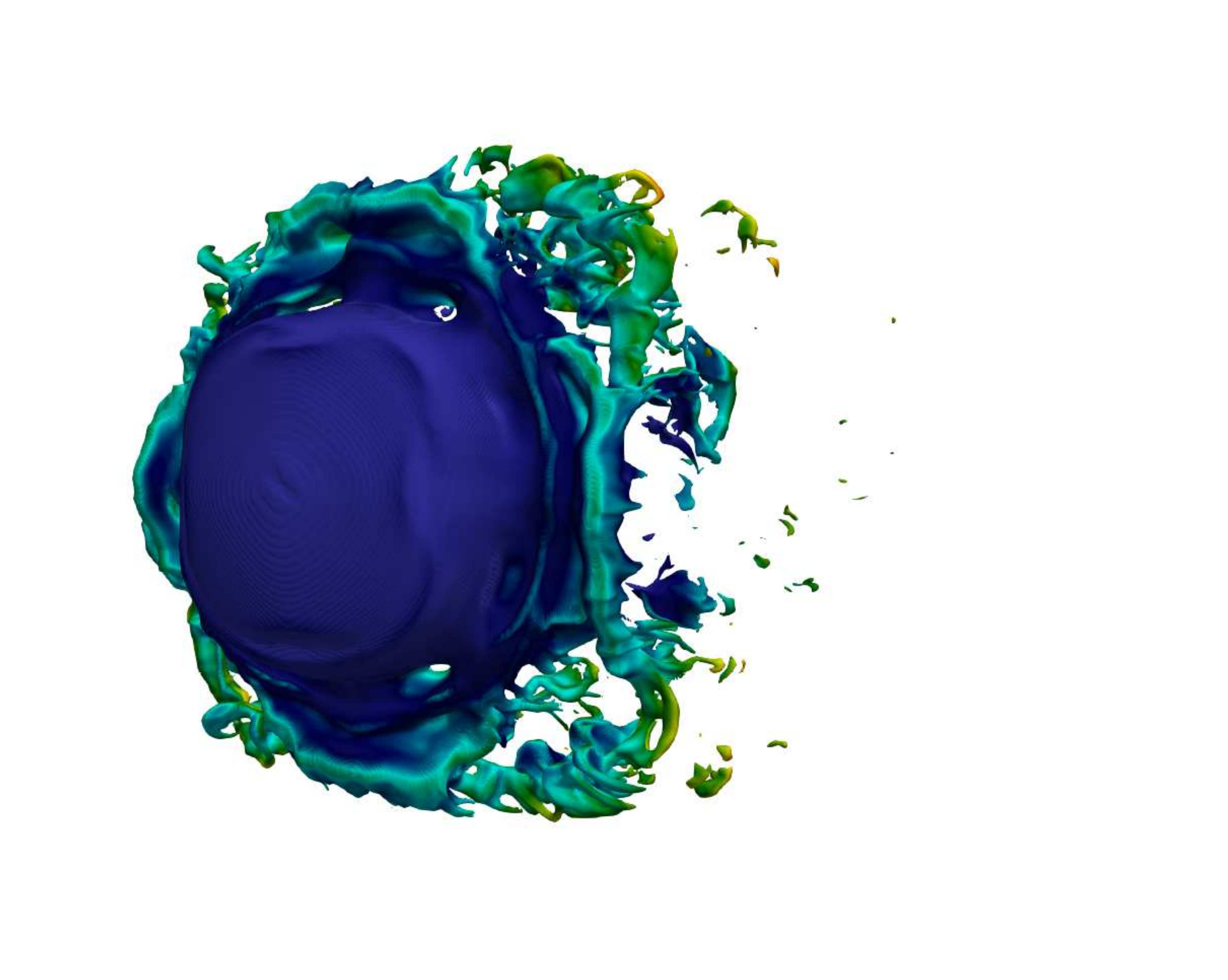}
\caption{Second breakup at $\tau= 0.94$ for $\rm{We}=470$. }
\label{fig:st_f232}
\end{subfigure}
\begin{subfigure}[t]{0.45\textwidth}
\includegraphics[clip, trim=20cm 2cm 5cm 2cm,width=\textwidth]{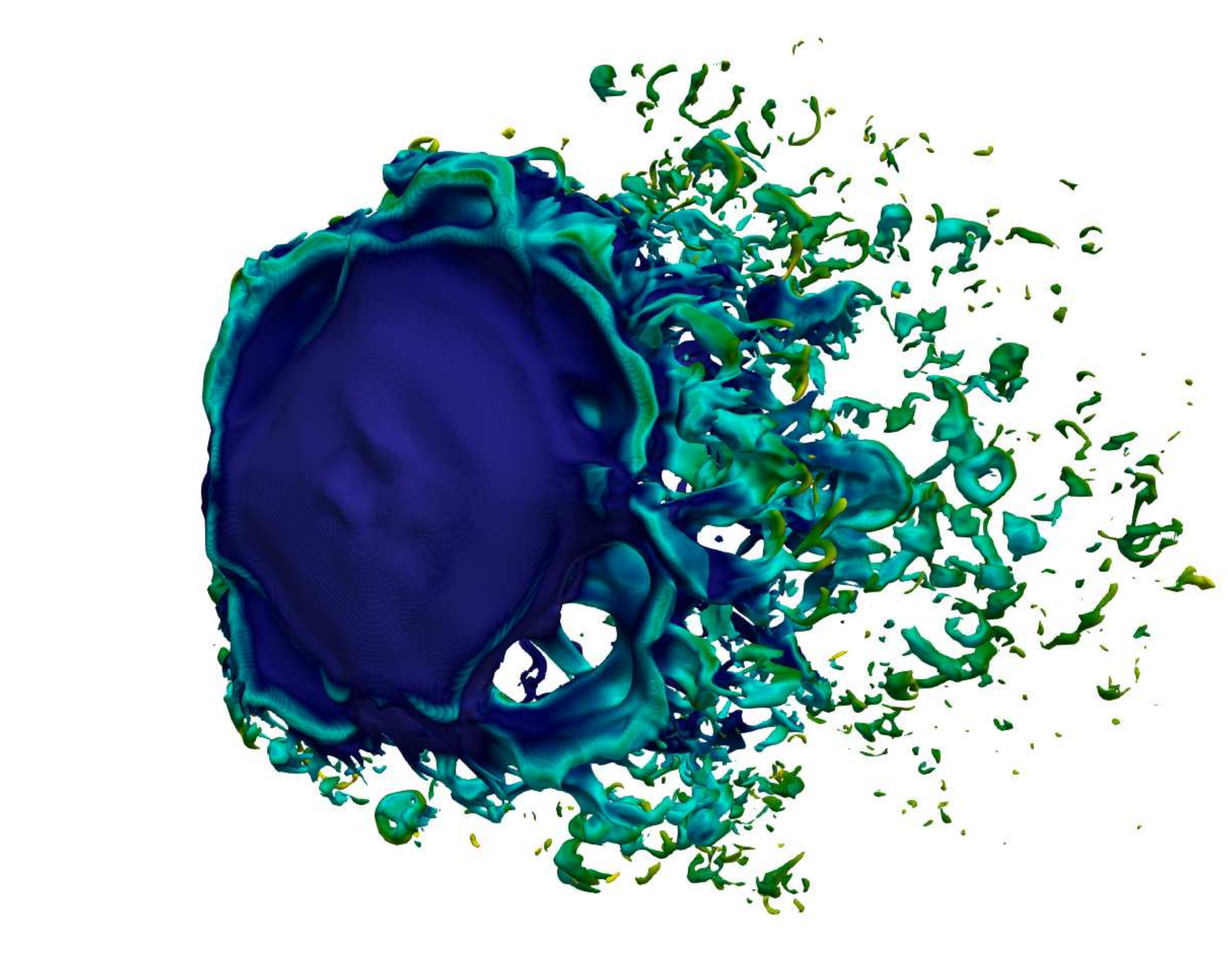}
\caption{Third breakup at $\tau= 1.33$ for $\rm{We}=470$. }
\label{fig:st_f329}
\end{subfigure}
\\
\begin{subfigure}[t]{0.45\textwidth}
\includegraphics[clip, trim=20cm 2cm 5cm 2cm,width=\textwidth]{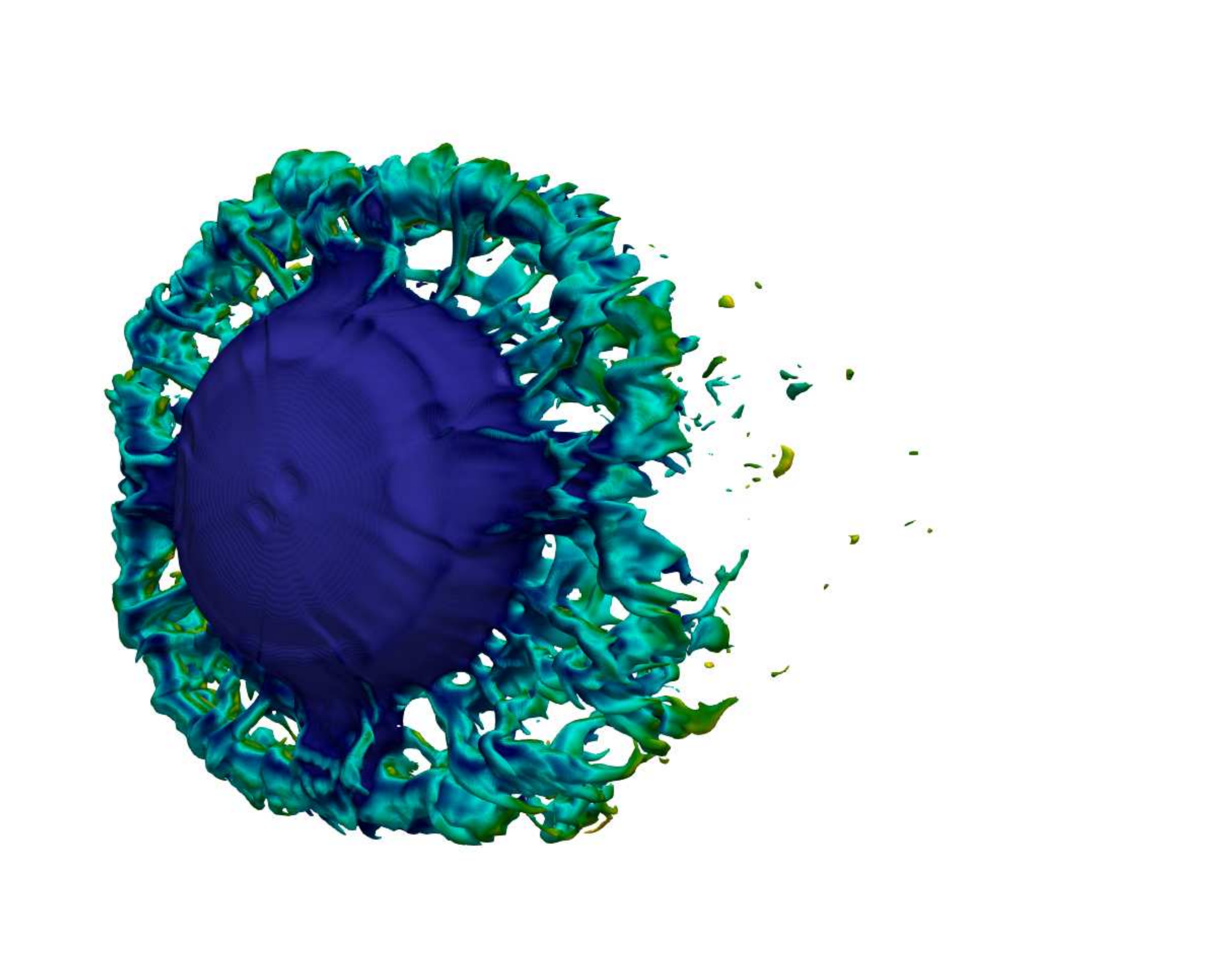}
\caption{Second breakup at $\tau= 0.91$ for $\bWe$. }
\label{fig:nost_f225}
\end{subfigure}
\begin{subfigure}[t]{0.45\textwidth}
\includegraphics[clip, trim=20cm 2cm 5cm 2cm,width=\textwidth]{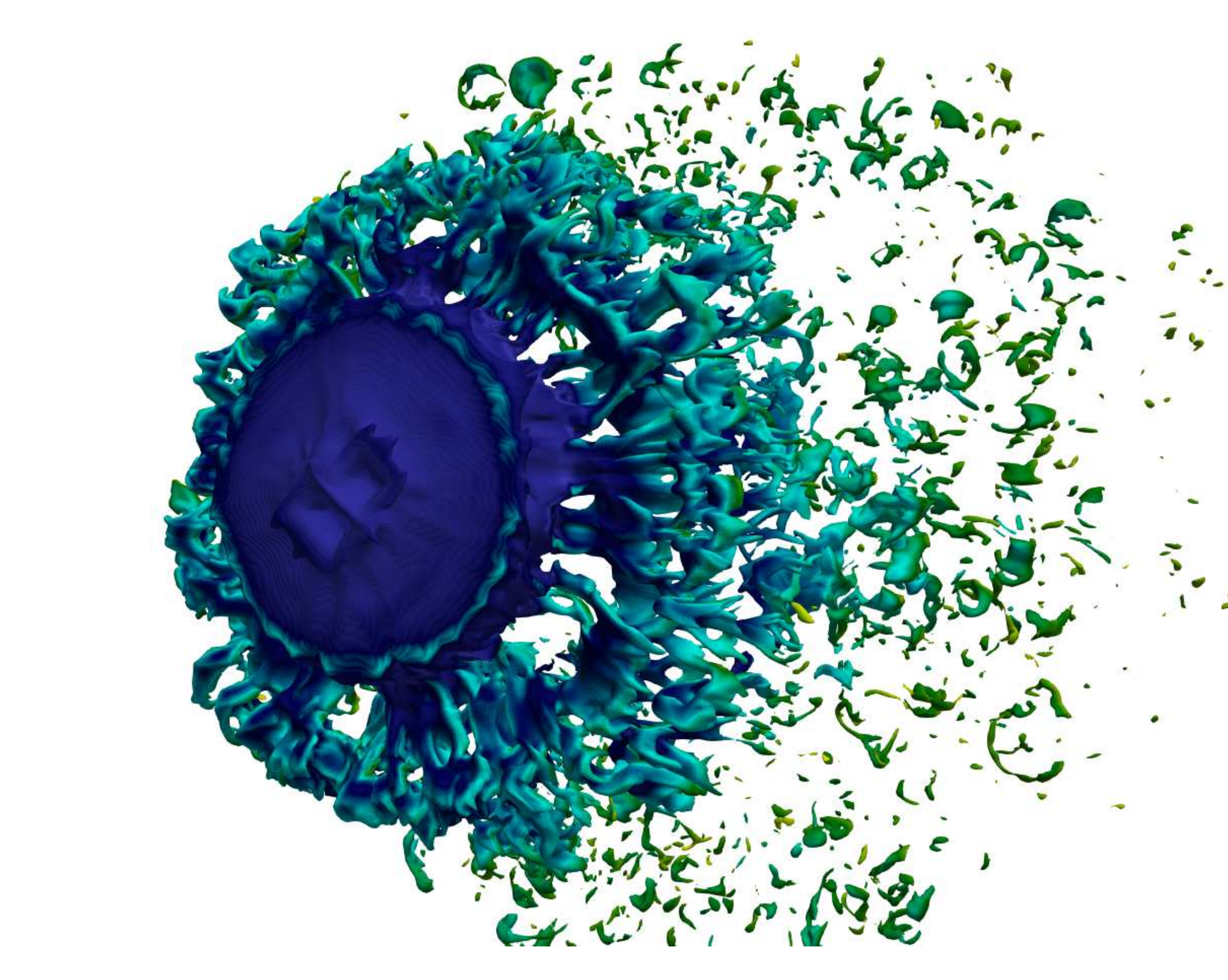}
\caption{Third breakup at $\tau= 1.25$ for $\bWe$. }
\label{fig:nost_f309}
\end{subfigure}
\caption{Recurrent breakups for both simulations at $\rm{We}=470$ and $\bWe$.}
\label{fig:recurrentBreakupSnapshots}
\end{figure}
While the stages of the breakup and the formation of the ligaments are well
described by the above mechanisms, our simulations and experiments suggest that the pattern of events occurs repeatably at different scale throughout the course of the breakup.  
In particular, when considering the breakup for a Weber number of $\rm{We}=470$ as shown in figures
\ref{fig:st_f232}-\ref{fig:st_f329}, we can qualitatively observe a very
similar droplet morphology and breakup behavior for times $\tau=0.94$ and
$\tau=1.33$. 
Hence, the process repeats after the initial breakup or ligament formation
process.  An analogous process can also be observed in our simulations for $\bWe$ as demonstrated in figures
\ref{fig:nost_f309}-\ref{fig:nost_f225}, where the breakups are observed 
at $\tau=0.91$ and $\tau=1.25$. 
This suggests a negligible effect of capillary forces on such recurrent
shedding behavior.

\begin{figure}
  \centerline{\includegraphics[width=\textwidth]{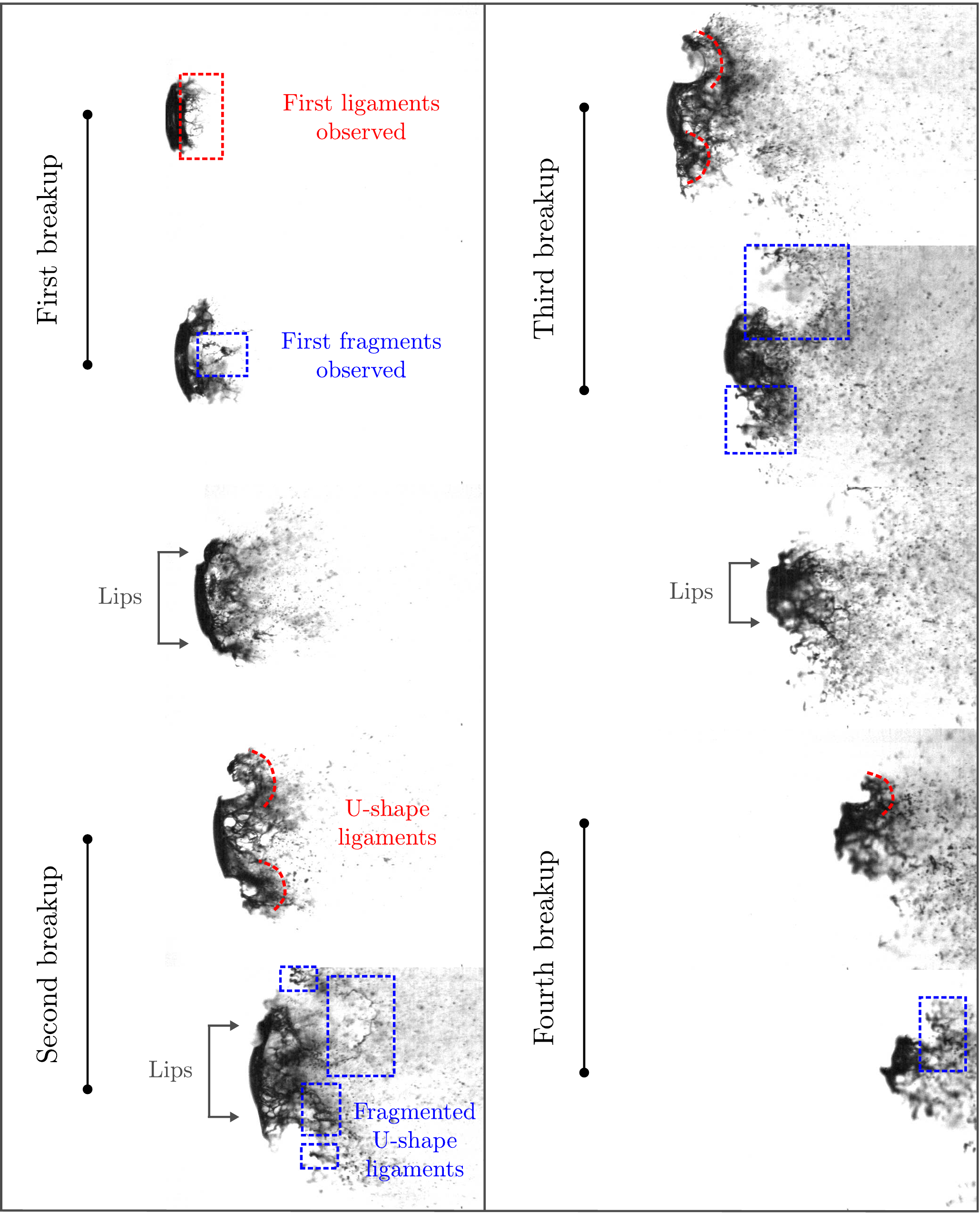}}
  \caption{Experimental snapshots of the recurrent breakup mechanism for a typical case at $\rm{We}=295$.
   }
  \label{fig:recurrentBreakupSnapshots_exp}
\end{figure}

Consulting our experiments, we can investigate this effect  for
a longer time span than possible with the numerical simulations.  To that end,
the snapshots are post-processed manually and the breakup times are defined as
the time between subsequent snapshots in which the ligaments are still attached
to the main droplet core and when they have been shed from the droplet. This
has been done for 29 independent experiments with varying Weber numbers.
Starting from the formation of the first ligaments, we can observe a total of
four breakups until the droplet is entirely disintegrated. 
A typical sequence of four breakups as captured by the experiments is  
shown in figure \ref{fig:recurrentBreakupSnapshots_exp}. 
Clearly visible are the ligaments and their subsequent fragmentation.
%

\begin{figure}
\centering
\begin{subfigure}[b]{0.45\textwidth}
    \includegraphics[width=\textwidth]{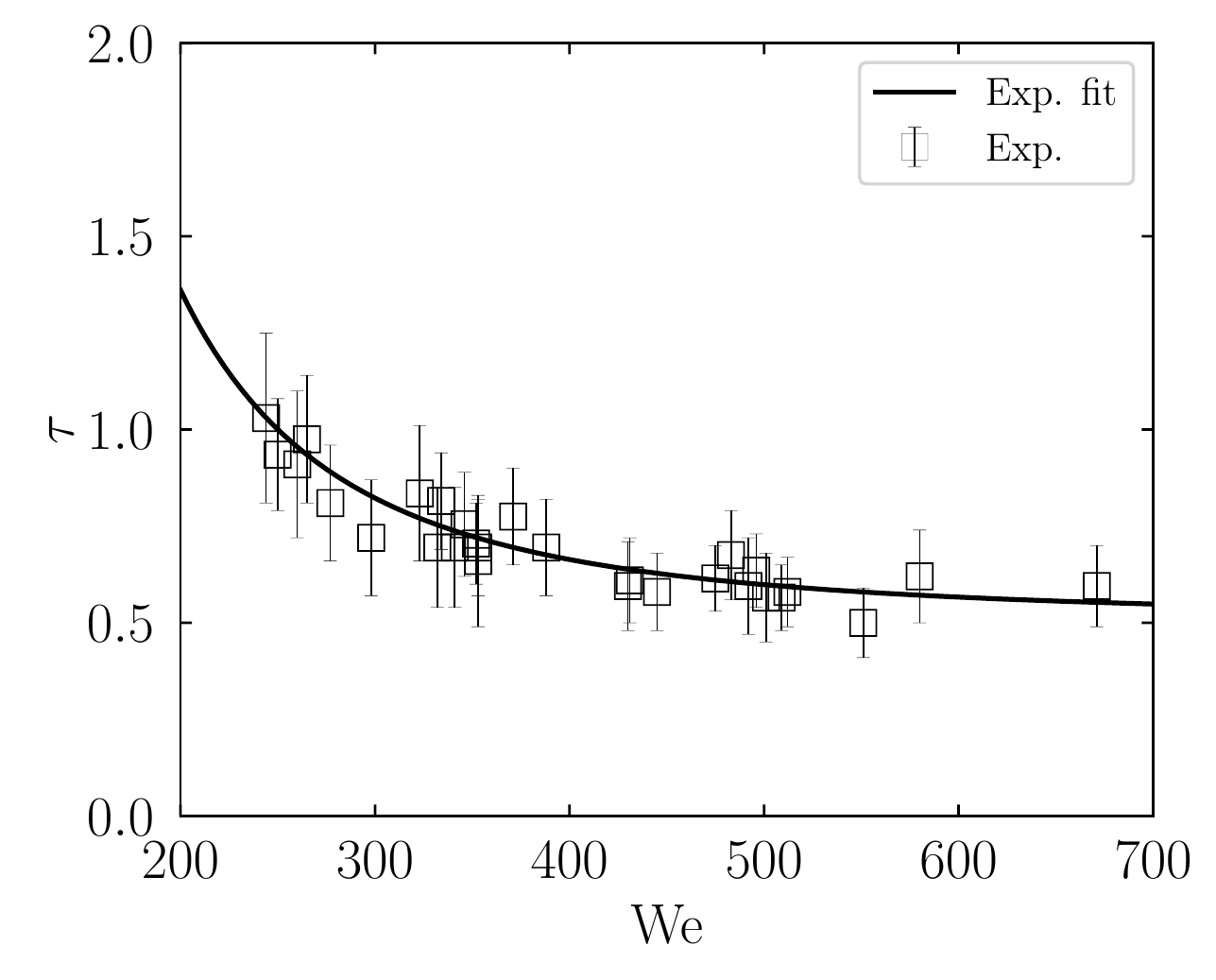}
    \caption{First breakup.}
    \label{fig:breakup1}
\end{subfigure}
\begin{subfigure}[b]{0.45\textwidth}
    \includegraphics[width=\textwidth]{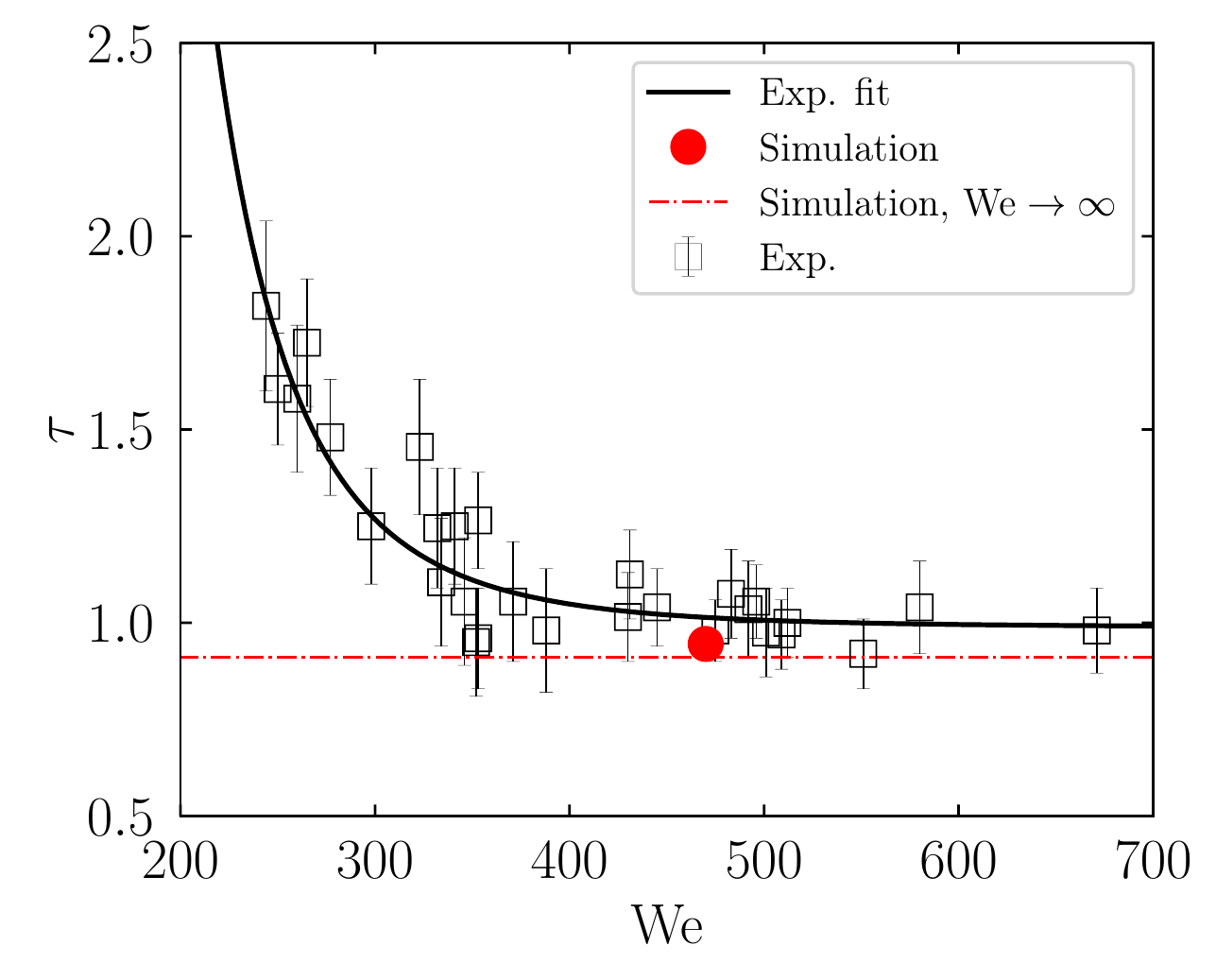}
    \caption{Second breakup. }
    \label{fig:breakup2}
\end{subfigure}
\\
\begin{subfigure}[b]{0.45\textwidth}
    \includegraphics[width=\textwidth]{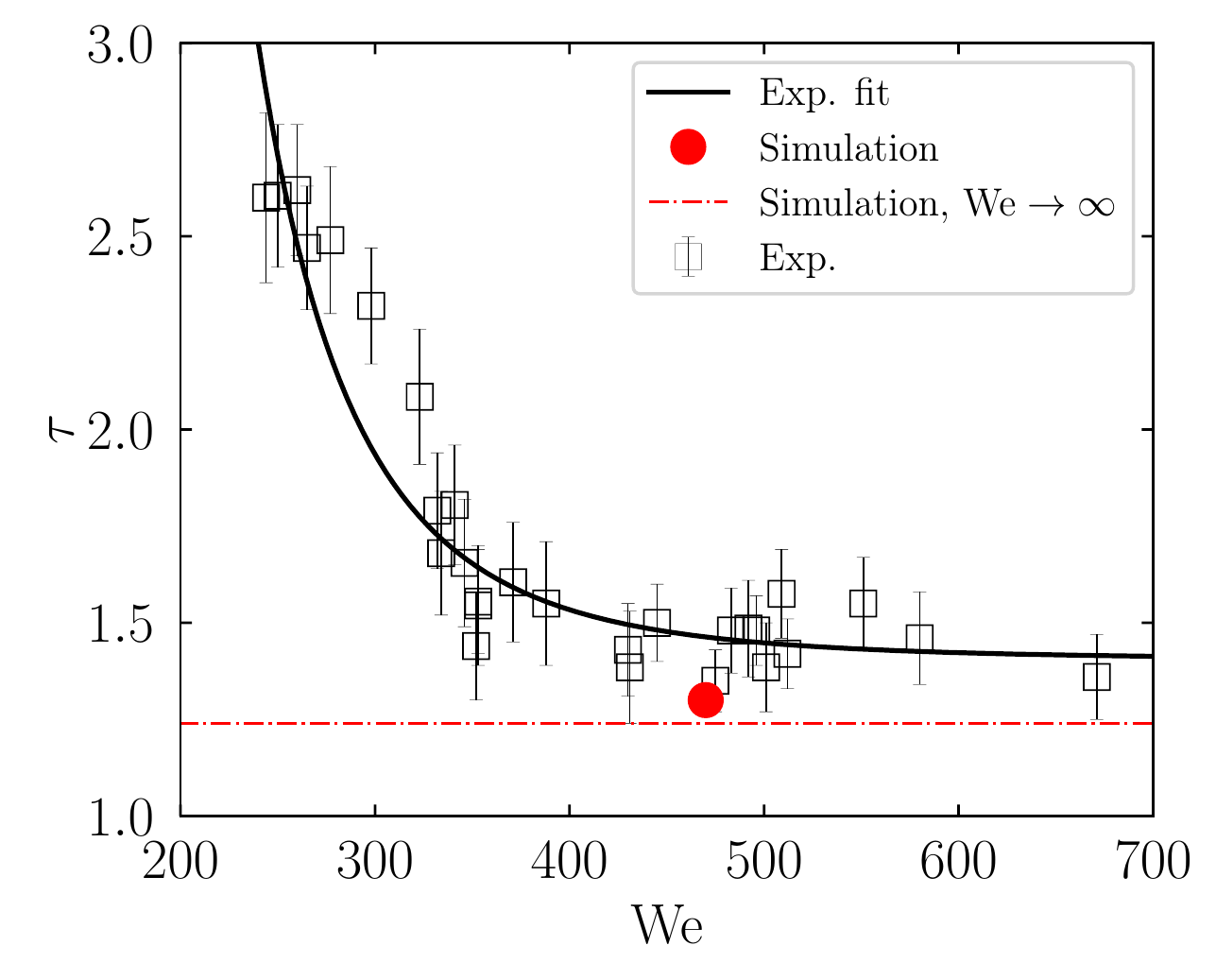}
    \caption{Third breakup. }
    \label{fig:breakup3}
\end{subfigure}
\begin{subfigure}[b]{0.45\textwidth}
    \includegraphics[width=\textwidth]{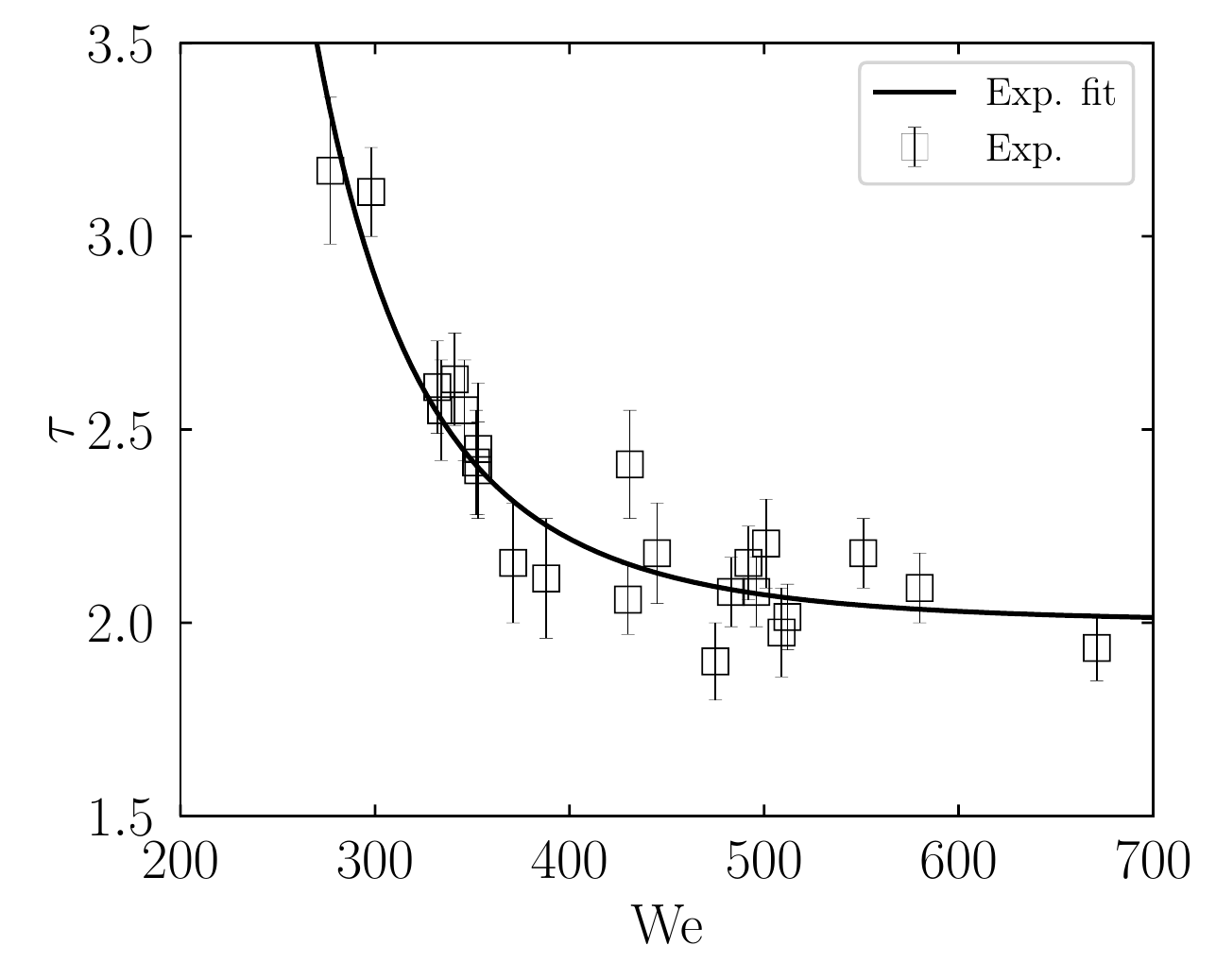}
    \caption{Fourth breakup. }
    \label{fig:breakup4}
\end{subfigure}
\caption{Breakup time dependence on the Weber number.  The error bars indicate
the time difference between subsequent snapshots before and after the breakup
of the ligaments.}
\label{fig:BreakuptimeVsWeber}
\end{figure}

Recording the breakup times as a function of the Weber number (figure \ref{fig:BreakuptimeVsWeber}) reveals that, while 4 breakups are recorded for all Weber numbers, capillary effects have an impact on the breakup times, reaching a nearly constant value beyond $We > 400$.
The error bars indicate the time difference between subsequent snapshots before
and after the breakup of the ligaments.
%
%
%
A comparison with our simulations in figure \ref{fig:BreakuptimeVsWeber} shows that the first and second breakup times in the simulation agree
well with the experimental observations for the second and third breakup for both the We=$470$ and $\bWe$ simulations.
This is believed to be due to the fine-scale nature of the first breakup, which our simulations are not able to resolve.
Nonetheless, this further validates both the simulations as well 
as the post-processing methodology for the experiments.

%

\begin{figure}
  \centerline{\includegraphics[width=0.7\textwidth]{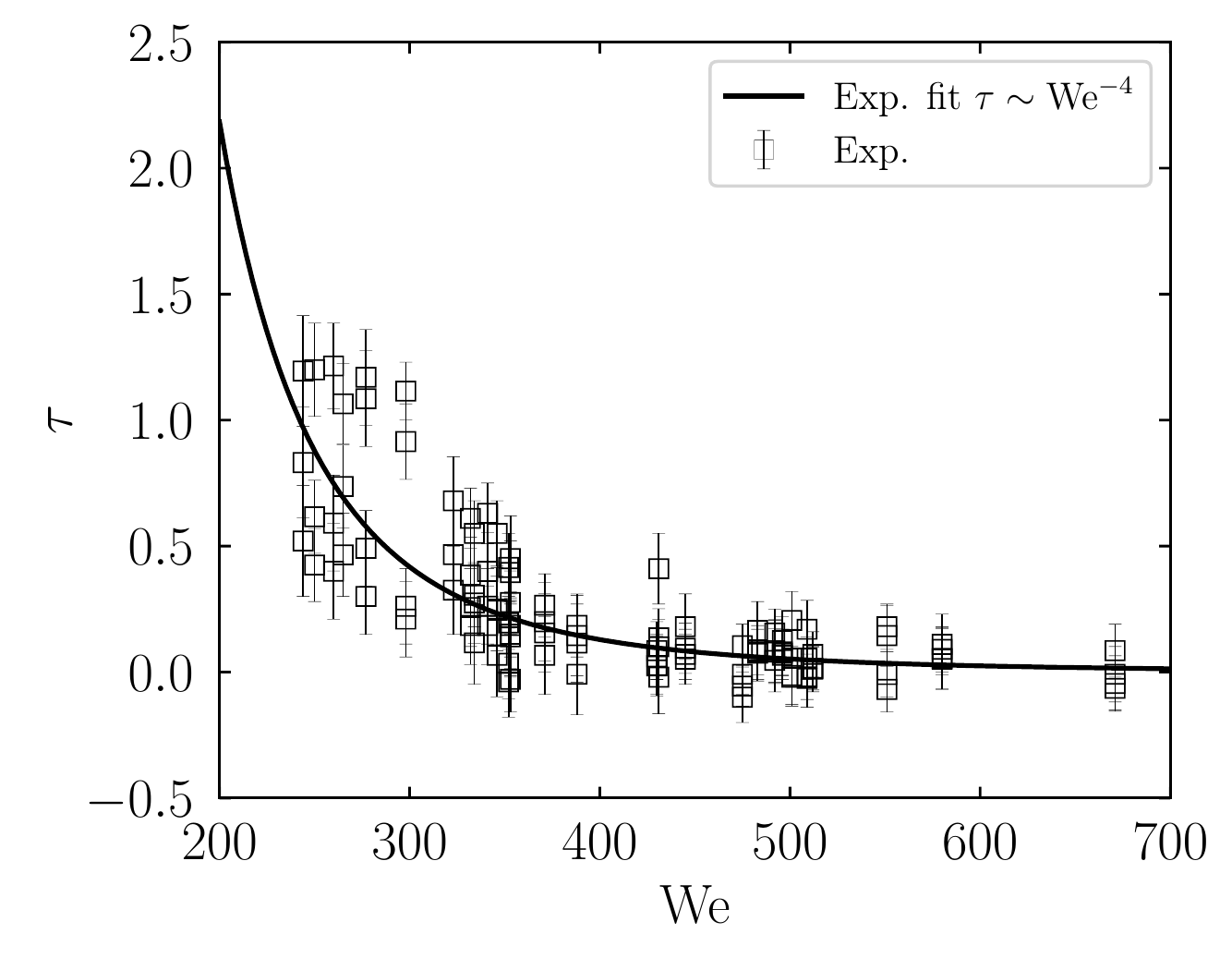}}
  \caption{Collapse of all breakup times when shifted by their asymptotic breakup time. The least-square fit suggests $\tau \sim {\rm{We}}^{-4}$.}
  \label{fig:breakupAll}
\end{figure}

The functional dependence of the breakup onset 
to the Weber number is estimated using a nonlinear least-square fit
of the experimental data of the form $\tau \approx a {\rm{We}}^b+c$ and plotted 
alongside the measurements in figure \ref{fig:BreakuptimeVsWeber}.
The fitting coefficients are reported in table \ref{tab:fitCoeffs}, which reveal
a similar functional dependence for all breakup times.
In particular, the similar exponents $a$ and constant difference between the offsets $c$ for all breakups reveal
an approximately constant 
frequency between the breakups. This suggests that only the initial onset of the 
breakup is a function of the Weber number, whereas the breakup and its frequency 
are independent of capillary effects.
Put differently, surface-tension seems to delay the onset of the initial ligament shedding but does not affect the frequency of the recurrent shedding.
Hence, these curves can be collapsed when shifted by $c$, which is reported 
in figure \ref{fig:breakupAll}. 
The fit suggests that the breakup onset scales with $\tau \sim {\rm{We}}^{-4}$.
\begin{table}
\centering
\begin{tabular}{l|c|c|c}
Breakup No. & $a$              & $b$    & $c$           \\
\hline
$1$         & $4.07\cdot 10^5$  & $-2.47$ & $0.51$      \\
$2$         & $5.22\cdot 10^{12}$ & $-5.36$ & $0.99$    \\
$3$         & $8.58\cdot 10^{11}$ & $-4.93$ & $1.41$    \\
$4$         & $1.34\cdot 10^{12}$ & $-4.92$ & $2.00$    \\
$all$       & $5.41\cdot 10^{9}$  & $-4.08$ & $0$       \\
\hline
\end{tabular}
\caption{Fitting coefficients for a nonlinear least-square fit of the form $\tau \approx a {\rm{We}}^b+c$. The confidence interval for all fitting coefficients is 0.95.}
\label{tab:fitCoeffs}
\end{table}

We conjecture that the breakup frequency is dominated 
by aerodynamical effects only.
Such effects are dominant in the initial stages of the droplet deformation 
and previous studies of aerobreakup  have
shown a similar drag coefficient to that of a flow past a sphere (see, e.g., \cite{meng2018numerical}).
In our case, it is instructive to evaluate the Strouhal number 
$St=\frac{fD}{U}$ for the observed breakups.
However, the characteristic length-scale, associated to the 
deforming and disintegrating droplet and its wake, is a priori not clear.
\begin{figure}
  \centerline{\includegraphics[width=0.7\textwidth]{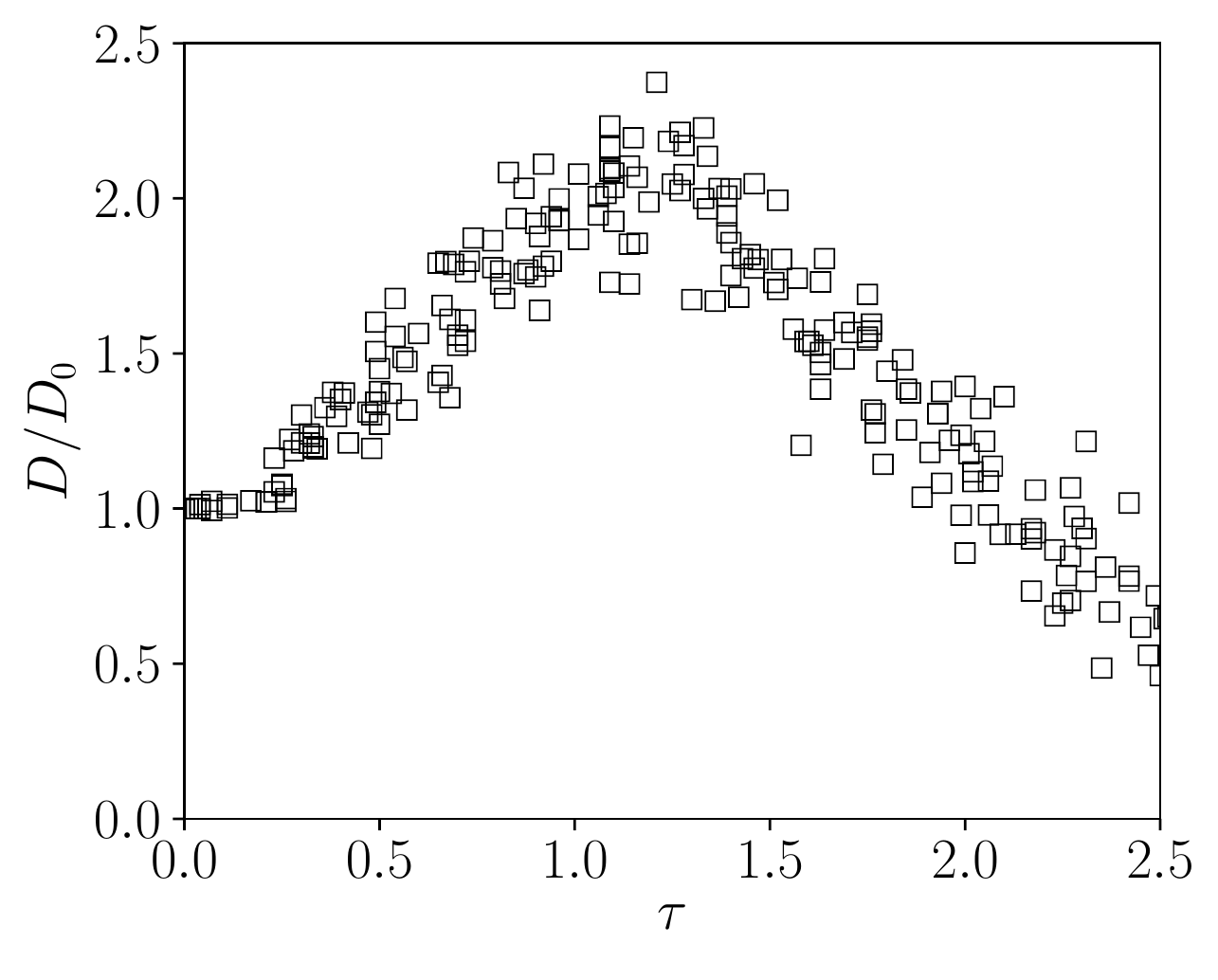}}
  \caption{Diameter evolution for Weber numbers in the range $\rm{We}=[200, 700]$.}
  \label{fig:diameterEvolution}
\end{figure}
To that end, in figure \ref{fig:diameterEvolution}, we plot the evolution of
the droplet core diameter throughout the breakup process for all
experimental runs with Weber numbers in the range of $\rm{We}=[200, 700]$.  
The diameter is measured directly from the shadowgraphs and defined as the maximum extent of the coherent liquid body of the droplet.
Remarkably, the diameter evolution is independent of the Weber
number and the droplet expands up to twice the size of its initial diameter
for all experiments.
Note that at later stages for $\tau>1$, the core diameter decreases due to shedding and $D<2D_0$ but the small-scale structures and mist shed
form the droplet, yield an effective diameter which is larger than the diameter of the droplet core as plotted on figure \ref{fig:diameterEvolution}.
Hence, the maximum droplet diameter $D_m= 2D_0$ as a characteristic length scale
seems to be the natural choice for reduction of the breakup frequency.
This yields a mean Strouhal number of $\rm{St}\approx0.217$ when using all
experimental data.
Using the fitted data, one similarly obtains $\rm{St}\approx0.18$.
Both frequencies agree well with what is observed for the flow past a rigid sphere
(see, e.g., \cite{achenbach1974vortex,kim1988observations,Sakamoto1990Vortex}).
This suggests that the recurrent breakup is indeed induced by classical 
vortex shedding.



\section{Conclusion}
Three dimensional simulations and high-magnification shadowgraphy visualizations of the aerobreakup of 
a water droplet have been performed to capture the underlying mechanisms leading the ligament formation and disintegration.
The numerical simulations are first validated with respect to experimental results by comparison of
observed deformation and the evolution of the center of mass of the droplet, and the number of ligaments that are formed during breakup.  
An analysis of the perturbations arising on the liquid sheet surrounding the droplet 
shows good qualitative agreement with the concept of transverse azimuthal modulation. 
From the numerical results, modes associated with the transverse destabilization have been found by 
means of an azimuthal Fourier decomposition of the flow field. 
These correspond to the wavenumber of ligaments which form subsequent to the initial growth of azimuthal modulation. 
Finally, we experimentally and numerically show what we believe to be the first observation of recurrent shedding of ligaments. 
The first breakup event occurs at a time that depends weakly on the Weber number and with stronger capillarity the breakup is delayed, whereas subsequent events occur at the same fixed time interval, independent of We.  The frequency of recurrence of breakup events is therefore driven primarily by inertia.  
By casting the frequency as a Strouhal number based on the pancaked droplet diameter, which reached a value of $2D_0$ compared to the initial droplet diameter, independent of We, we find $St \approx 0.2$, which supports a hypothesis that this shedding behavior is related to vortex shedding in the wake of the deformed droplet.


\section*{Acknowledgements}
The experimental work (LBP and HER) was supported by the R\'egion Nouvelle-Aquitaine as part of the SEIGLE project (2017-1R50115) and the CPER FEDER project. 
BD acknowledges support from the Swiss National Science Foundation Grant No. P2EZP2\_178436.
Computations associated to parallel performance have utilized the Extreme Science and Engineering Discovery Environment, which is supported by the National Science Foundation grant number CTS120005.

\bibliography{references.bib}
\bibliographystyle{jfm}

\end{document}